\documentclass[
superscriptaddress,
12pt,
tightenlines,
amsmath,amssymb,amsthm,
aps,
prx,
]{revtex4-2}

\usepackage{soul}

\usepackage{graphicx}
\usepackage{dcolumn}
\usepackage{bm}
\usepackage{hyperref}
\usepackage[mathlines]{lineno}
\linenumbers\relax 
\usepackage[utf8]{inputenc}

\usepackage{braket}
\usepackage{physics}
\usepackage{mathtools}
\usepackage{tikz-cd}
\usepackage{dsfont}
\usepackage{scalerel}
\usepackage{ulem}
\usepackage{cancel}

\usepackage{mathbbol}
\DeclareSymbolFontAlphabet{\amsmathbb}{AMSb}%

\newcommand{\tp}{\intercal}
\def\ie{\textit{i.e., }}
\def\eg{\textit{e.g., }}

\def\N{\mathcal{N}}
\def\opL{\mathcal{L}}
\def\Z{\mathcal{Z}}
\def\W{\mathcal{W}}

\def\Rset{\mathds{R}}
\def\O{\mathbf{O}}
\def\X{\mathbf{X}}
\def\x{\mathbf{x}}

\def\l{\mathbf{l}}
\def\bSigma{\boldsymbol{\Sigma}}
\def\bmu{\boldsymbol{\mu}}
\def\bxi{\boldsymbol{\xi}}
\def\bseta{\boldsymbol{\eta}}
\def\AA{\mathbb{A}}
\def\CC{\mathbb{C}}
\def\DD{\mathbb{D}}

\def\GM{\mathbb{\Gamma}}
\def\HH{\mathbb{H}}
\def\BB{\mathbb{B}}
\def\II{\mathbb{1}}
\def\JJ{\mathbb{J}}
\def\VV{\mathbb{V}}
\def\MM{\mathbb{M}}
\def\QQ{\mathbb{Q}}
\def\VVt{\mathbb{V}_t}
\def\UU{\mathbb{U}}
\def\UUt{\mathbb{U}_t}
\def\Ssigma{\mathbb{\Sigma}}
\def\vF{{\bf F}}
\newcommand{\JJJ}{\mathsf{J}}

\newcommand{\Thet}{\mathbb{\Theta}}
\def\hrho{\hat{\rho}}
\def\hrhog{\hat{\sigma}}
\def\hA{\hat{A}}
\def\hB{\hat{B}}
\def\hC{\hat{C}}
\def\hH{\hat{H}}
\def\hHH{\hat{\HH}}
\def\hMM{\hat{\MM}}
\def\hL{\hat{L}}
\def\hO{\hat{O}}
\def\hT{\hat{T}}
\def\hU{\hat{U}}
\def\hq{\hat{q}}
\def\hp{\hat{p}}
\def\hx{\hat{x}}
\def\hy{\hat{y}}
\def\huno{\hat{1}}
\def\hbx{\hat{\mathbf{x}}}
\def\hby{\hat{\mathbf{y}}}

\newcommand{\ese}{\end{subequations}}

\begin{document}

\nolinenumbers

\title{Fluctuation and dissipation in memoryless open quantum evolutions}

\author{Fabricio  Toscano}   
\email{toscano@if.ufrj.br \ (corresponding author) }
\affiliation{Instituto  de F\'isica,  Universidade  Federal do  Rio de  Janeiro,
  21941-972, Rio de Janeiro, Brazil}

\author{Gustavo M. Bosyk}
\email{gbosyk@gmail.com}
\affiliation{Instituto  de  F\'isica La  Plata,  CONICET--UNLP,  1900 La  Plata,
  Argentina}
\affiliation{Universit\`a degli Studi di Cagliari, Cagliari, Italy}

\author{Steeve Zozor}
\email{steeve.zozor@cnrs.fr}
\affiliation{Universit\'e Grenoble  Alpes, CNRS, Grenoble INP,  GIPSA-Lab, 38000
  Grenoble, France}

\author{Mariela Portesi}
\email{portesi@fisica.unlp.edu.ar}
\affiliation{Instituto  de  F\'isica La  Plata,  CONICET--UNLP,  1900 La  Plata,
  Argentina}

\date{\today}

\begin{abstract}
Von Neumann  entropy rate for  open quantum systems  is, in general,  written in
terms of entropy production and entropy  flow rates, encompassing the second law
of  thermodynamics.  When  the  open-quantum-system  evolution  corresponds to  a
quantum  dynamical semigroup,  we find a  decomposition  of the  infinitesimal
generator of  the dynamics, that allows  to relate the von  Neumann entropy rate
with the divergence-based  quantum Fisher information, at any  time.  Applied to
quantum Gaussian channels that are dynamical semigroups, our decomposition leads
to the  quantum analog of  the generalized  classical de Bruijn  identity, thus expressing 
the quantum fluctuation--dissipation relation in that kind of
channels.  Finally, from this perspective, we analyze how stationarity arises.
\end{abstract}

\maketitle

\section{Introduction}

The study on quantum communication  channels that describe the input-output maps
corresponding  to  quantum mechanical  operations  is  at  the core  of  quantum
information theory (see \eg\cite{Holevo2019}).   Formally, these maps are given,
in  the  Schrödinger  picture,  by  completely  positive  and  trace  preserving
operations acting on  density operators, whereas, in the  Heisenberg picture, by
identity-preserving operations acting on  observables.  A physically interesting
property of these maps is that their composition 
is also a quantum
channel.  Accordingly, the set of quantum channels forms a semigroup.  Moreover,
quantum channels have an inverse only when they describe a unitary evolution.  A
unitary  quantum channel  requires for  its implementation a quantum system completely  isolated    
from  its   surrounding   environment.    In  practice,   the
implementation of unitary channels is  a great challenge. Consequently, the most
common  situation corresponds  to non-unitary  quantum channels  that lead  to a
degradation  of  information  during their use.   To  be  able  to  assess  the
degradation effects, it is important  to describe the evolutionary trajectory of
the  system that implements the channel.  The  description is  possible through  continuous maps  over time.
This is  the point where quantum  information theory meets open  quantum systems
theory.

We focus  in this paper  on quantum channels  that are members  of one-parameter
semigroups,  for which  the density  operator $\hrho_t$  of the  evolved quantum
system    is     solution    of    a    time-independent     Markovian    master
equation~\cite{Heinosaari2010}.   The  set of  these  channels  forms a  quantum
dynamical    semigroup   in    the    case   of    finite-dimensional    Hilbert
spaces~\cite{Lindblad1976}  as well  as  in  the case  of  Gaussian channels  in
continuous-variable systems~\cite{Demoen1979,  Heinosaari2010, Tarasov2008}, \ie
channels  in  infinite-dimensional  separable Hilbert  spaces.   These  quantum
dynamical semigroups are  suitable to describe memoryless  quantum channels that
are  continuous  in  time~\cite{Caruso2014},  specially  in  continuous-variable
systems~\cite{Eisert2007, Heinosaari2010}.

In general, an  open system dynamics, given by the interaction  
between the system and
its environment,  is modeled by a  combination of both deterministic  and random
effects.  One of the main approaches  to the systematization of these random and
deterministic effects  lies in  the so  called fluctuation-dissipation relations,
both   in    the   classical~\cite{Einstein1905,    Ornstein1927,   Nyquist1928,
  Onsager1931I, Onsager1931II} and  quantum domains~\cite{Callen1951, Green1952,
  Green1954,        Kubo1957,        Bernard1959,       Efremov1969}        (see
also~\cite{Stratonovich1992}).   Roughly  speaking,  these  are  relations  that
connect  the  deterministic characteristic  of  a  system with  its  fluctuating
aspects, both in the equilibrium  and non-equilibrium regimes.  Examples of these
relations  are   the  linear   and  non-linear   Markov  fluctuation-dissipation
relations~\cite{Stratonovich1992}.  In classical  theory, these relations appear
within Markov processes  where the probability distribution of  the variables of
interest satisfies a  Fokker--Planck equation.  In particular, in the  case of a
linear   Fokker--Planck  equation   the  corresponding   fluctuation-dissipation
relations are just a set of  identities that link
the correlations between the
variables of  interest with the  intensities of the fluctuation  and dissipation
effects in a stationary regime~\cite{Stratonovich1992}.

On the other  hand, the ubiquitous notion of entropy  is transversal to physical
and information theories, since it arises naturally as a measure of uncertainty,
randomness or lack of information about the state of a system~\cite{Balian2004}.
For a  quantum system described  by a density  operator $\hrho$, the  von Neumann
entropy is  defined by $S[\hrho] =  - \Tr(\hrho \ln \hrho)$  where $\Tr$ denotes
the trace of an operator (see  \eg~\cite{Wehrl1978}). 
This quantity plays a key
role in different contexts as in the study of entropy production in open quantum
systems  in  general (see  \eg~\cite{Landi2020})  and  of quantum  communication
channels where different entropic quantities  serve to characterize 
the
information-processing performance of  the channels (see \eg~\cite{Holevo2012, Ohya2004},
among others).  Here, we  are interested in the rate of
change of the von Neumann entropy $S[\hrho_t]$ in one-parameter semigroups.  The
entropic  analysis of its temporal behavior 
is  of  crucial importance,  for
instance, in  order to improve the  communication rates of quantum  channels, as
well as for studying stationary situations.

In open  quantum theory,  the study  of the  rate of  change of  $S[\hrho_t]$ is
usually oriented by a thermodynamic perspective~\cite{Spohn1978, Landi2020}.  In
this framework the focus is on the  decomposition of the rate of change into two
parts: one corresponding to the so called entropy production rate $\Pi_t$, and the
other one to the entropy flux rate $\dot\Phi_t$.  From this point of view, the second law
of  thermodynamics is  nothing but  the statement  $\Pi_t\geq 0$.   Considerable
progress has been made in recent years  from this perspective in particular in a
general    definition    of    $\Pi_t$    for    general    quantum    processes
(see~\cite{Landi2020} and references therein).  However, those approaches do not
focus on the characterization of the  fluctuation and dissipation aspects of the
dynamics.

At this point a natural question  arises: is it possible to identify generically
random and deterministic effects in the master equation for the density operator
and for rate  of change of the von Neumann  entropy in one-parameter semigroups?
Here  we provide  a  positive answer  for  all finite-dimensional  one-parameter
quantum channels and for all infinite-dimensional Gaussian one-parameter quantum
channels.  Both situations are described by  Lindblad master equations whose solutions form
quantum  dynamical  semigroups~\cite{Lindblad1976,  Demoen1979,  Heinosaari2010,
  Tarasov2008,  BreuerPetruccione2002,   AlickiFannes2001}.   Our   approach  is
inspired by a well-established result in classical information theory known
as de  Bruijn identity~\cite{Shannon1948,Stam1959} which quantifies  the rate of
change  of the  Shannon entropy  of a  classical random  variable, output  of an
additive  Gaussian  noise  channel.   More precisely, this   
identity
quantifies  how  fast   the  channel  becomes  more  random  in   terms  of  the
non-parametric Fisher  information.  The  fact that a 
de Bruijn  identity can
also be formulated  for quantum systems was first  posed in~\cite{Konig2014} for
the quantum diffusion  process, a useful result for  formulating quantum entropy
power  inequalities~\cite{Konig2013,  Konig2014,   DePalma2014}.   However,  the
quantum diffusion  process is a  special subclass of the  Gaussian one-parameter
quantum channels that does not have dissipation effects.  Therefore, our framework
is also  inspired by the  recently established  generalization of the de Bruijn
identity  for more  general classical  channels  modeled  by Langevin  forces described  by
stochastic  differential  equations~\cite{Toranzo2017, Wibisono2017},  capturing
the trade-off between the diffusion and dissipation terms of the evolution.

First, we propose a decomposition of the non-unitary infinitesimal generator for
any quantum  dynamical semigroup  into two  terms.  The first  one allows  us to
relate the time derivative of the  von Neumann entropy with the divergence-based
quantum Fisher  information~\cite{Nagaoka2005}.  Accordingly, this term  will be
associated with the fluctuations due to the noise introduced by the open quantum
dynamics.  In addition,  the second term will be connected  with the dissipative
contributions of the dynamics.

Afterwards, 
we focus on Gaussian channels, which are among the most important channels
in   information    processing   in   both   classical~\cite{Shannon1948,
  CoverThomas2006}   and   quantum~\cite{Holevo2007,   Caruso2008,   Eisert2007,
  Weedbrook2012,  Adesso2014} domains,  as  they provide  a  faithful model  for
attenuation  and noise  effects in  most communication  schemes, modeled  by  
linear Lindblad master equations~\cite{Heinosaari2010}.   For these channels, we
find  the fully  quantum  counterpart of  the generalization  of  the de  Bruijn
identity for Fokker--Planck channels~\cite{Toranzo2017, Wibisono2017}.  In these
cases, our proposal captures in a clear manner the diffusion and dissipation contributions to the rate of  change of  the  von Neumann  entropy  given by  the  open quantum  dynamics.
Finally, we analyze stationary situations within our framework.

The paper is organized as follows.  In Sec.~\ref{subsec:deBruijnA} we review the
classical de Bruijn  identity for channels with additive Gaussian  noise and its
generalization     for     multidimensional classical   channels     corresponding     to
Ornstein--Uhlenbeck  processes, \ie  modeled  by  Fokker--Planck equations  with
linear  drift and  constant diffusion.   In Sec.~\ref{subsec:deBruijnB}  we show
that the diffusion  term in the  generalized classical de  Bruijn identity can be
written  in terms  of the  Fisher information  of the  probability distribution  
under  the  action  of  Langevin  forces.   This  provides  a  novel
interpretation of  the diffusion term  as a measure  of the noise  introduced by
these forces.  We also show that the second term in the generalized classical de
Bruijn identity corresponds to the average flux of the dissipative forces, \ie a
measure of the change of the probability distribution in the directions opposite
to these  forces.  In  Sec.~\ref{sec:vNopensystemsA} we  present our  first main
result, namely  a decomposition of  the non-unitary infinitesimal  generator for
any  quantum dynamical  semigroup that  splits the  dynamics into  two different
parts: fluctuation and dissipation.  Section~\ref{sec:vNopensystemsB}  is devoted
to the presentation of the notion of divergence-based quantum Fisher information
and some  of its  properties.  In  Sec.~\ref{sec:vNopensystemsC} we  present our
second   main  contribution,    namely   the   application   of    the   results   of
Sec.~\ref{sec:vNopensystemsA} in order to  obtain a  closed formula  for the  von Neumann
entropy rate  of change for quantum  dynamical semigroups, where we discriminate 
the contributions of fluctuation  and dissipation.  In Sec.~\ref{sec:QdBiGaussianA},
we present the basic formalism used to describe one-parameter Gaussian channels.
In      Sec.~\ref{Sec:QuandBiGDS},     we      apply    our      results     of
Sec.~\ref{sec:vNopensystemsC} to obtain the quantum
counterpart  of the  generalized classical  de Bruijn  identity in the case of one-parameter  Gaussian
channels.   We also  show  that  the diffusion  and  dissipation  terms of  the identity obtained,   
admit  an   interpretation   completely  analogous   to  that   given
in Sec.~\ref{subsec:deBruijnB} for the corresponding terms  in the generalized classical de Bruijn identity.   In Sec.~\ref{Sec:QuandBiGDSGau}  we  specialize  the quantum de Bruijn
identity for evolving Gaussian states and in Sec.~\ref{subsec:StatStat} we study
the stationarity conditions  in one-parameter Gaussian channels in  light of our
formalism.  We  conclude with Sec.~\ref{sec:conclusions} where  we summarize our
findings.  For ease of reading, some  auxiliary calculations and technical proofs are
presented in Appendices~\ref{appA} and~\ref{appB}.

\section{Rate of Shannon entropy  for a linear Fokker-Planck equation}
\label{sec:deBruijn}

\subsection{Generalized de Bruijn identity}
\label{subsec:deBruijnA}

In   classical  information   theory~\cite{Shannon1948,  CoverThomas2006},   the
additive Gaussian noise  is probably the most used noise  model to describe many
``natural'' random processes viewed  as a large sum of noise  sources and due to
the    central   limit    theorem~\cite{BrockwellDavis1987,   AshDoleanDade1999,
  CoverThomas2006}.   The Gaussian  noise channel  corresponds to  $X_t =  X_0 +
\sqrt{t} \, Z$, being $Z \sim \N(0,1)$  a normally distributed random variable with
zero mean  and unit  variance, independent  of the  input random  variable $X_0$
(that admits a  finite variance).  Here, $t  > 0$ is a  parameter (usually time)
that controls the amount of randomness  added to the system.  The evolution with
respect to $t$ of the probability density function $p_t(x)$ of the output random
variable  $X_t$  is  ruled  by  the  heat  equation  $\dv{p_t(x)}{t}  =  \frac12
\pdv[2]{p_t(x)}{x}$.   The  de Bruijn  identity  plays  a fundamental  role  in
classical information theory~\cite{Stam1959, CoverThomas2006, Barron1986}, as it
quantifies the rate  of change of the Shannon entropy $h[p_t]  = - \int_\Rset p_t(x)
\ln p_t(x) \dd  x$ at the output  of the channel, in terms of  
the Fisher information
$J[p_t] = \int_\Rset \left( \pdv{\ln p_t(x)}{x} \right)^2 \, p_t(x) \, \dd x = -
\int_\Rset  \pdv[2]{\ln  p_t(x)}{x}  \,   p_t(x)  \,  \dd  x$~\footnote{$J$  is
generally  called   nonparametric  corresponding  to  usual   parametric  Fisher
information  when  $p(x,\theta)=q(x-\theta)$,  so that  differentiating  in  $x$
replaces differentiating  in $\theta$,  therefore $\theta$  drops from  $J$.  In
addition,  the definition  holds  provided  that $p$  is  differentiable in  the
closure of its support.}.  More precisely,
\begin{equation}
\label{eq:dBi}
\dv{h[p_t]}{t}= \frac12  J[p_t].
\end{equation}
Because $J  > 0$, the  fundamental interpretation of  this equality is that $X_t$ becomes more and more  random as $t$ (time) increases, with $J$ quantifying
how  fast.   Applications  of  de  Bruijn  identity  include  derivation of 
relevant information-theoretical inequalities~\cite{Blachman1965,  Dembo1991, Barron1986,
  Guo2005},  definition of  
an   entropic  temperature~\cite{Narayanan2012},  proof of 
monotonicity of some  statistical complexity measures~\cite{Rudnicki2016}, among
others~\cite{Park2012}.

De  Bruijn identity extends to  the multivariate case~\cite{CoverThomas2006,
  Toranzo2017},  and its form has  recently  been obtained  for  slightly  more  general
channels  modeled  by  Langevin  forces  described  by  stochastic  differential
equations~\cite{Toranzo2017, Wibisono2017}.  A similar expression for the rate of
change  of  Shannon  entropy was previously  given in  the  thermodynamic
context~\cite{Daems1999, Bag2001},  but without pointing  out its link  with 
Fisher  information.    In  particular,  consider  a   multidimensional  channel
corresponding   to   an   Ornstein--Uhlenbeck  process. It is   characterized   by   an
$N$-dimensional  random  vector  $\X_t$  that follows  a  stochastic  differential
equation with  linear drift  $\bmu(\X_t ,  t) =  \AA \X_t  - \bxi$  and constant
diffusion parameter $\Ssigma(\X_t , t) = \Ssigma$,
\begin{equation}
\label{eq:OU}
\dd \X_t = \left( \AA \X_t - \bxi \right) \dd t + \Ssigma \, \dd \mathbf{W}_t.
\end{equation}
Here $\AA$ and $\Ssigma$  are real constant matrices with dimensions  $N\times N$ and
$N\times M$ ($M\leq N$) respectively, being $\Ssigma$ of full rank, $\bxi$ is an
$N$-dimensional constant  real vector, and $\mathbf{W}_t$  is an $M$-dimensional
standard    Wiener   process~\cite{Uhlenbeck1930,    Risken1996,   Gardiner2004,
  Vatiwutipong2019}.  For $\bmu = 0$ (null vector),  the $N-$dimensional Gaussian additive channel is recovered.
Note  that the  drift corresponds  to  the deterministic  effects, whereas  the
diffusion term $\Ssigma \, \dd \mathbf{W}_t$ characterizes the random contributions
in the dynamics.

The  probability  density  function  of the  random  variable  $\X_t$  satisfies the 
Fokker--Planck equation~\cite{Risken1996, Gardiner2004, Vatiwutipong2019}:
\begin{equation}
\label{eq:FPE}
\dv{p_t(\x)}{t}  = -  \pdv{\x^\tp} \left(  (\AA{\x}  - \bxi)  p_t(\x) \right)  +
\frac12 \pdv{\x^\tp} \DD \pdv{\x} p_t(\x),
\end{equation}
where $\pdv*{\x} = \left( \pdv*{x_1} ,  \ldots , \pdv*{x_N} \right)^\tp $ is the
gradient with  respect to  $\x =  (x_1, \ldots,  x_N)^\tp $  and $\DD  = \Ssigma
\Ssigma^\tp$  is  the   diffusion  matrix,  which  is   symmetric  and  positive
semidefinite of rank $M$.    
For this  channel, the generalized  de Bruijn
identity is~\cite{Wibisono2017}
\begin{equation}
\label{eq:BruGen}
\dv{h[p_t]}{t} = \frac12 \tr\left( \DD \JJ[p_t] \right) + \tr\left( \AA \right),
\end{equation}
where $h[p_t]  = - \int_{\Rset^N}  p_t(\x) \ln p_t(\x) \dd^N \x $ \ ($\dd^N \x = \dd x_1 \ldots \dd x_N$) is the Shannon entropy, $\tr$ denotes trace of matrices, 
and the Fisher information matrix is defined as
\begin{equation*}
\JJ[p_t]  = \int_{\Rset^N}  \pdv{\ln p_t(\x)}{\x}  \pdv{\ln p_t(\x)}{\x^\tp}  \,
p_t(\x) \, \dd^N \x
\end{equation*}
which,    under    regularity    conditions,    can   be    written    in    the
form~\cite{CoverThomas2006}
\begin{equation}
 \label{claFishermat}
 \JJ[p_t]   =   -
\int_{\Rset^N} p_t(\x)\,\HH[\ln p_t(\x) ]\, \dd^N \x,
\end{equation}
with  $\HH  =  \pdv{\x}  \pdv{\x^\tp}$   the  Hessian  matrix  operator.   The
multivariate  de Bruijn  identity  is  recovered when  $\AA  =  0$ (null  matrix) and $\bxi=0$
(null vector), whereas  the
original (scalar)  de Bruijn  identity is recovered  when in addition   $N=1, \ \DD=\II_1$ . 
For a continuous variable system with an underlying symplectic structure,
$N = 2  n$ with $n$ the  number of classical modes described  by the canonically
conjugate variables of position $x_i = q_i$ and momentum $x_{n+i} = p_i$, \ $i =
1 , \ldots , n$.  Typical examples  of this type are the description of Brownian
motion and the electric field in a laser where $q_i$ and $p_i$ are the classical
quadratures of the electric field~\cite{Risken1996, Carmichael1999}.

\subsection{Fluctuation-dissipation relation in the generalized de Bruijn identity}
\label{subsec:deBruijnB}

Let us  rewrite the  generalized de Bruijn  identity~\eqref{eq:BruGen} so  as to
highlight  the   contributions  of  the  fluctuations   and  dissipation.   More
precisely,  by expressing  the Ornstein--Uhlenbeck  process as  $\dv*{\X_t}{t} =
\vF_{dr} + \vF_f$, where  $\vF_{dr} = \AA \X_t - \bxi = \bmu(\X_t  , t)$ are the
drift forces and $\vF_f =  \Ssigma \dv*{\mathbf{W}_t}{t}$ are the fluctuating or
Langevin  forces, the  first term in the  r.h.s.\  of~\eqref{eq:BruGen} can be
explicitly identified as induced by fluctuations due to the Langevin forces,
whereas the second one by dissipation due to the drift forces.

To this end, let us first rewrite the diffusion matrix under the form
\begin{equation}
\label{DDdecomposition}
\DD = \sum_{m=1}^M \bSigma_m \bSigma_m^\tp,
\end{equation}
where  $\bSigma_m$ are  the column  vectors of  $\Ssigma$. From the
linearity  of  the  trace and the relation  $\tr(  \bSigma_m  \bSigma_m^\tp  \JJ  )  =
 \bSigma_m^\tp  \, \JJ  \,  \bSigma_m$~, the  first  term in  the
r.h.s.\ of~\eqref{eq:BruGen} can therefore be expressed as
\begin{equation}
\label{Flucontribution}
 \frac12    \tr\left(   \DD    \JJ[p_t]   \right)    =   \frac12    \sum_{m=1}^M
 \left. J[p^{(\theta , \bSigma_m)} ; \theta] \right|_{\theta=t},
\end{equation}
where $p^{(\theta  , \bSigma_m)}  = p_t(\x  - \delta\theta \, \bSigma_m )$  is the
translated probability  distribution ($\delta\theta =  \theta - t$),  and indeed
$\bSigma_m^\tp \, \JJ[p_t]  \, \bSigma_m = \left.  J[p^{(\theta  , \bSigma_m)} ;
  \theta]  \right|_{\theta =  t}$  is  the Fisher  information  $J[p; \theta]  =
\int_{\Rset^N}  p \,  ( \pdv*{\ln  p}{\theta} )^2  \dd^N \x$ for the  translated
distribution.

We  remark  that  the  r.h.s.\   of~\eqref{Flucontribution}  makes  explicit  the
dependence on the Langevin forces, $\bSigma_m$, and  also on the rank $M$ of the
diffusion matrix $\DD$.
From the unicity  of $\DD = \Ssigma  \Ssigma^\tp$, $\Ssigma$ is unique  up to an
orthogonal transformation, \ie $\DD  = \bar{\Ssigma} \bar{\Ssigma}^\tp = \Ssigma
\Ssigma^\tp  $ if and only if $\bar{\Ssigma} =  \Ssigma \mathbb{Q}$  with $\mathbb{Q}$  an
arbitrary $M \times M$ orthogonal matrix,  such a transform is equivalent to use
any other\sout{s} Langevin forces, $\bar{\bSigma}_m$,  which are the columns vectors of
$\bar{\Ssigma}$. Furthermore,  this also holds for  any $M \times  \bar{M}$ matrix
$\mathbb{Q}$ of the Stiefel manifold ($\bar{M} \ge M$), \ie such that $\mathbb{Q} \,
\mathbb{Q}^\tp$ is the $M \times M$  identity, so that the stochastic differential equation~\eqref{eq:OU} with $\bar{\Ssigma}$ is
the equivalent Langevin equation of that with the full rank $\Ssigma$. This shows
that we can also express each particular matrix $\DD$ with an expression 
like in \eqref{Flucontribution} but with the $\bar{M}$ ($\bar{M}> N$) columns  of a matrix $\bar{\Ssigma}$.
However,  the Langevin forces turn out to be not all independent. This kind of 
situation arises 
for the diffusion matrices $\DD$ that come from linear Lindblad master equations (see Eq.\eqref{DDdecomposition2}).

Moreover, from  the expression  $D_{KL}[p^{(\theta ,  \bSigma_m)} \|  p_t(\x)] =
\frac12  \left.   J[p^{(\theta ,  \bSigma_m)}  ;  \theta] \right|_{\theta  =  t}
\delta\theta^2  + o(\delta\theta^2)$,  where $D_{KL}$  is the  Kullback--Leibler
divergence, or relative  entropy~\cite{Kullback1968, CoverThomas2006}, the first
term in the r.h.s. of the  classical de Bruijn identity~\eqref{eq:BruGen}  is essentially a
measure  of  the  noise  induced  by the  Langevin  forces  $\bSigma_m$  on  the
probability distribution $p_t(\x)$.

Secondly,  from the  definition  of  $\vF_{dr}(\x) =  \AA  \x  - \bxi$,  its
  divergence is given by $\pdv{\vF_{dr}(\x)}{\x^\tp}  = \tr \left( \AA \right)$,
  which  gives, together  with $\expval{\pdv{\vF_{dr}(\x)}{\x^\tp}}_{\!   p_t} =
  \int_{\Rset^N} \pdv{\vF_{dr}(\x)}{\x^\tp } \; \; p_t(\x) \; \dd^N \x$,
  \begin{equation}
  \label{trAinterpretedFlux}
  \tr\left( \AA \right) =  \expval{\pdv{\vF_{dr}(\x)}{\x^\tp }}_{\! p_t}.
  \end{equation}
  From  the fact  that $p_t(\x)  \; \vF_{dr}(\x)$  vanishes in  the boundary  of
  $\Rset^N$ ($X_t$ admits a mean), so  that the integral of $\pdv{\left( p_t(\x)
    \; \vF_{dr}(\x)\right)}{\x^\tp}  = \pdv{\vF_{dr}(\x)}{\x^\tp } \;  p_t(\x) +
  \pdv{p_t(\x)}{\x^\tp} \, \vF_{dr}(\x)$ vanishes, one equivalently has 
  \begin{equation}
  \label{trAinterpretedChange}
  \tr\left(  \AA   \right)  =  -  \int_{\Rset^N}   \;  \pdv{p_t(\x)}{\x^\tp}  \,
  \vF_{dr}(\x) \; \dd^N \x . 
  \end{equation}
  Note that  this result can  be recovered  directly from the  expression of
  $\vF_{dr}(\x)$, together with the identity
  \begin{equation}
  \label{identity}
  \int_{\Rset^N}   \; \x  \pdv{p_t(\x)}{\x^\tp}  \; \dd^N \x = - \II .
  \end{equation}
  This is because  $\x \, p_t(\x)  $ vanishes in  the boundary of  $\Rset^N$ ($X_t$
  admits  a  mean),  so  that  the   integral  of  $\pdv{\left(  \x  \;  p_t(\x)
    \right)}{\x^\tp} = \II \, p_t(\x) + \x \, \pdv{p_t(\x)}{\x^\tp } $
  vanishes.  
Therefore, the quantity $\tr\left( \AA \right)$  can be interpreted as minus the
total  amount  of  change  of  the probability  distribution  $p_t(\x)$  in  the
directions given by the linear drift  force $\vF_{dr}(\x)$, or as the average of
the drift force flux~\cite{Daems1999}.  In  addition, by expressing $\AA = \AA_S
+ \AA_{AS}$ with $\AA_S = (\AA + \AA^\tp) / 2$ and $\AA_{AS} = (\AA - \AA^\tp) /
2$, we have $\tr\left( \AA \right)  = \tr\left( \AA_S \right)$.  Accordingly, we
rewrite  the  drift  force   as  $\vF_{dr}(\x)=  \vF_S(\x)+\vF_{AS}(\x)$  with
$\vF_S(\x) = \AA_S\, \x$ and $\vF_{AS}(\x) = \AA_{AS} \x - \bxi$ , in order to
highlight that  only the  symmetric part  contributes to  the drift  force flux.
Moreover, whenever  $\tr(\AA) <  0$, we associate  $\vF_S$ with  a dissipative
force  and $\vF_{AS}$  with a  non-dissipative one.   These names  are justified
because  in   the  context  where  the   Fokker--Planck  equation~\eqref{eq:FPE}
describes a mechanical system, $\vF_S$ represents the force in the phase space
of the  system that does not  conserve the energy and  $\vF_{AS}$ corresponds to
the Hamiltonian phase space force that conserves the energy.

Finally,  we  collect  the  expressions  given in  Eqs.~\eqref{Flucontribution}
and~\eqref{trAinterpretedFlux} to obtain
\begin{equation}
\label{eq:BruGenR}
\dv{h[p_t]}{t}  =  \frac12  \sum_{m=1}^M  \left. J\left[ p^{(\theta  ,  \bSigma_m)}  ;
  \theta \right] \right|_{\theta = t} + \expval{\pdv{\vF_{dr}(\x)}{\x^\tp}}_{\! p_t} .
\end{equation}
Let  us  observe   that  the  first  term   in  the  r.h.s  of   the  de  Bruijn
identity~\eqref{eq:BruGen}  (or   equivalently~\eqref{eq:BruGenR})  is  strictly
positive,  because each  Fisher information  is positive,  \ie $\left.   J\left[
  p^{(\theta , \bSigma_m)}  ; \theta \right] \right|_{\theta = t}  > 0$.  On the
other hand, the  second term has no  definite sign due to its  dependence on the
sum of  the eigenvalues  of $\AA_S$,  which can be  negative, positive  or zero.
Therefore, a necessary  condition for the existence of a  stationary solution of
the Fokker--Planck equation~\eqref{eq:FPE} is the entropic balance between these
quantities.   This can  happen only  if~\eqref{trAinterpretedFlux} is  negative,
which  leads to  a condition  on the  eigenvalues of  $\AA_S$.  Recall  that the
existence  of  a stationary  solution  also  requires  the  matrix $\AA$  to  be
asymptotically stable, that is, all  its eigenvalues must strictly have negative real
part.  In principle, this  condition  has not a direct connection with respect
to the one on the eigenvalues of $\AA_S$.

\

In the  sequel, we will obtain  an analogous generalized de  Bruijn identity for
the rate of change of the von  Neumann entropy in Gaussian channels whose Wigner
function satisfies  the Fokker--Planck equation~\eqref{eq:FPE}. Before  that, in the next section we will show that such a fluctuation-dissipation  relation in
the quantum case has its origin in a particular decomposition of the non-unitary
infinitesimal generator of the evolution in quantum dynamical semigroups.

\section{Rate of von Neumann entropy for quantum dynamical semigroups}
\label{sec:vNopensystems}

\subsection{Decomposition of the infinitesimal generators of the evolution}
\label{sec:vNopensystemsA}

In the Schr\"odinger  picture, the dynamics of an open  quantum system with density operator $\hrho_t$
is   generally  modeled   by   a   Lindblad  master   equation   (LME)  of   the
form~\cite{Tarasov2008}
\begin{equation}
\label{eq:mastereq}
\dv{\hrho_t}{t}     =    \opL[\hrho_t]     =    \opL_{\mathrm{U}}[\hrho_t]     +
\opL_{\mathrm{NU}}[\hrho_t],
\end{equation}
where
\begin{equation}
\label{eq:LU}
\opL_{\mathrm{U}}[\hrho_t]  = \frac{1}{\imath  \hbar}[\hH,\hrho_t]
\end{equation}
and
\begin{equation}
\label{eq:LNU}
\opL_{\mathrm{NU}}[\hrho_t]  =  \frac{1}{2\hbar} \sum_{k=1}^K  \left( 2  \hL_k\hrho_t
\hL_k^\dag - \hL_k^\dag \hL_k \hrho_t  - \hrho_t \hL_k^\dag \hL_k \right)
\end{equation}
are  the infinitesimal  generators  of the  unitary  and non-unitary  evolution,
respectively, with $\hH$ the Hamiltonian of  the system and $\{\hL_k\}_1^K$ the Lindblad
operators.    When   $\opL$   is   time   independent,   the   formal   solution
of~\eqref{eq:mastereq}    is    $\hrho_t    =    e^{t\opL}    \hrho_0$,    where
$\{\Lambda_t=e^{t\opL}\}_{t\geq0}$ is  a quantum dynamical semigroup  (QDS).  At
least   for  finite   dimension,  any   QDS   is  precisely   described  by   a
LME~\cite{Lindblad1976}.  Here, $\opL$ could be bounded, or unbounded as happens
for QDSs in Gaussian channels~\cite{Demoen1979,Heinosaari2010,Tarasov2008}.

In   the   Heisenberg  picture,   observables   $\hO_t$   evolve  according   to
$\dv{\hO_t}{t}           =            \bar{\opL}_{\mathrm{U}}[\hO_t]           +
\bar{\opL}_{\mathrm{NU}}[\hO_t]$,   where  $\bar{\opL}$   denotes  the   adjoint
superoperator of $\opL$~\cite{Tarasov2008}
~\footnote{Remind that, by definition,
$\langle \opL[\hat A], \hat B \rangle = \langle  \hat A , \bar{\opL}[\hat B] \rangle$ where 
$\langle \hat A , \hat B \rangle =  \Tr( \hat B^\dag \hat A )$ and the notation
``$^-$''  is  used to avoid confusion with  the  adjoint ``$\dag$''  of  an
operator.}.  

QDSs  are also called quantum  Markov semigroups~\cite{Carlen2017},
and   the    LME~\eqref{eq:mastereq}   is    also   called    Markovian   master
equation~\cite{Alicki2007}.  These denominations  emphasize the Markov semigroup
property:  $\Lambda_{t+s}=\Lambda_t\Lambda_s$~\cite{Tarasov2008},  which is  the
quantum version of an analogous semigroup property in time-homogeneous classical
Markov processes~\cite{Belton2019}.   Accordingly, in the same  way as classical
Markov processes describe  classical memoryless channels, QDSs  are suitable for
describing memoryless quantum channels.

\

We  propose  a general  decomposition  of  $\opL_{\mathrm{NU}}$, which,  in  the
context  of  Gaussian  channels,  encompasses   the  notions  of  diffusion  and
dissipation, as follows 
\begin{equation}
\label{L1L2L3}
\opL_{\mathrm{NU}} = \opL_1 + \opL_{2+3} , \qquad \opL_{2+3} = \opL_2+\opL_3,
\end{equation} 
with $\opL_1$, $\opL_2$ and $\opL_3$ real superoperators~\footnote{Remind that a
superoperator is real when $\opL(A)^\dag = \opL(A^\dag)$.} given by
\begin{subequations}
\label{eq:L1L2L3}
\begin{align}
\label{eq:L1}
\opL_1[\hO]  & =  -\frac{1}{2\hbar} \sum_{k=1}^K  ( [\hA_k  , [\hA_k  , \hO]]  +
    [\hB_k , [\hB_k ,\hO]]), \\
\label{eq:L2}
\opL_2[\hO] & = \frac{1}{2\hbar}  \sum_{k=1}^K \frac12 \{ [\hL_k ,\hL_k^\dag]
, \hO \}, \text{and} \\
\label{eq:L3}
\opL_3[\hO]  &  =  \frac{1}{2\hbar}  \sum_{k=1}^K  (\hL_k  \hO  \hL^\dag_k  -
\hL_k^\dag \hO \hL_k).
\end{align}
\end{subequations}
This follows from the cartesian decomposition of the Lindblad operators $\hL_k =
\hA_k + \imath \hB_k$ via the Hermitian operators $\hA_k$ and $\hB_k$ defined by
$\hA_k =  \frac12 (\hL_k +  \hL_k^\dag)$ and  $\hB_k = \frac1{2  \imath}(\hL_k -
\hL_k^\dag)$.

Notice that $\opL_1$  is selfadjoint, \ie $\bar{\opL}_1[\hO]  = \opL_1[\hO]$, as
is  $\opL_2$,  while  $\opL_3$  is antisymmetric,  \ie  $\bar{\opL}_3[\hO]  =  -
\opL_3[\hO]$.  For the unit operator, $\bar{\opL}_1[\huno] = \opL_1[\huno] = 0$,
$\bar{\opL}_2[\huno] =  \opL_2[\huno] =  \frac{1}{2\hbar} \sum_{k=1}^K  [\hL_k ,
  \hL_k^\dag]$ and $\bar{\opL}_3[\huno] = - \opL_3[\huno] = - \opL_2[\huno]$.
Consequently, $\opL_1$ and  $\opL_{2+3}$ are infinitesimal generators  of QDSs in
itself.  Moreover, $\opL_1$ is the generator of a unital QDS~\cite{Aniello2016}.

It  is   known~\cite{Tarasov2008} that   the  LME~\eqref{eq:mastereq}   is  invariant   under  the
transformations
\begin{subequations}
\label{Invtransf}
\begin{eqnarray}
\hH &\rightarrow&  \hH +  \hH^\prime \quad \left( \hH^\prime =  \frac{1}{2\imath \hbar}
\sum_{k=1}^K   (  \alpha_k^*   \hL_k  -   \alpha_k  \hL_k^\dag) \right),\quad\\
\hL_k &\rightarrow& \hL_k + \alpha_k \huno,
\end{eqnarray} \end{subequations}
for any set of complex numbers $\{\alpha_k\}_1^K$.  Under these transformations, the superoperators changes as follows 
\begin{subequations}
\label{Invinfgen}
\begin{eqnarray}
\opL_1 &\rightarrow&  \opL_1 \\
\opL_2 &\rightarrow&  \opL_2\;\;\mbox{and}\\
\opL_3    &   \rightarrow&    \opL_3   -    \frac{1}{\imath\hbar}[\hH^\prime   ,
  \cdot \;]. \label{Invinfgenc}
\end{eqnarray}
\end{subequations}

\subsection{Divergence-based quantum Fisher information}
\label{sec:vNopensystemsB}

As seen in Sec.~\ref{sec:deBruijn}, the  Fisher information is a key measure 
to quantify the rate  of change of the Shannon entropy  in the additive Gaussian
channel or in  the channel described by the Ornstein--Uhlenbeck  process, via the
classical  de   Bruijn  identities   Eqs.~\eqref{eq:dBi}  and~\eqref{eq:BruGen},
respectively~\cite{Stam1959, Wibisono2017}.
The  first attempt, to our knowledge,  to find  a  quantum counterpart  of~\eqref{eq:dBi} is  given
by K{\"o}nig and Smith in Ref.~\cite{Konig2014},  where  the  divergence-based  quantum  Fisher  information
(DQFI)  is related  to the  rate of  change  of the  von Neumann  entropy for an
evolution governed by  the quantum diffusion semigroup.  The DQFI  is one of the
forms of the Fisher  information  defined in  the  quantum  domain,  and is  precisely  that
appearing in the rate of von Neumann entropy in QDSs, in general.

The   DQFI   is   defined   as   the   second   derivative   of   the   relative
entropy~\cite{Nagaoka2005}, \ie
\begin{equation}
\label{eq:DQFI}
\left.  J_q\left[ \hrho_{\theta} ; \theta  \right] \right|_{\theta = \theta_0} =
\left. \dv[2]{S[\hrho_\theta \| \hrho_{\theta_0}]}{\theta}
\right|_{\theta = \theta_0},
\end{equation}
with $S[\hrho_\theta \| \hrho_{\theta_0}]  = \Tr(\hrho_\theta ( \ln \hrho_\theta
- \ln \hrho_{\theta_0}) )$.  
As noticed  in~\cite{Nagaoka2005}, this DQFI is  greater than the
quantum    Fisher    information    based    on    the    symmetric  logarithmic 
derivative~\cite{Helstrom1967,  Holevo1982, braunstein1994},  whereas the
respective classical versions coincide~\cite{CoverThomas2006}.

In  the particular  case of  the family  of density  operators $\hrho_{\theta  ,
  \hC_{\delta\theta}}  =   \hU_{\theta  -   \theta_0}  \,   \hrho_{\theta_0}  \,
\hU^\dag_{\theta  -   \theta_0}$,  generated  by  the   unitary  $\hU_{\theta  -
  \theta_0}$  with a  generator $\hC_{\delta\theta}  = \imath  \dv{\hU_{\theta -
    \theta_0}}{\theta}  \, \hU_{\theta  -  \theta_0}$ \  (where $\delta\theta  =
\theta - \theta_0$), we have~(see Appendix~\ref{appA})
\begin{equation}
\label{eq:QFIdef_commutator}
\left.     J_q[    \hrho_{\theta    ,    \hC_{\delta\theta}}    ;    \theta    ]
\right|_{\theta=\theta_0}   \!\!\!    =   \Tr\left(    \hrho_{\theta_0}   \left[
\hC_0 ,  \left[ \hC_{0} ,  \ln\hrho_{\theta_0} \right]
  \right] \right),
\end{equation}
where  $\hC_0=   \hC_{\delta\theta}|_{\theta=\theta_0}$.  It can be seen that when 
$\hC_{\delta\theta}$       is       independent      of       $\theta$,       the
expression~\eqref{eq:QFIdef_commutator}    reduces    to     the    one    given
in~\cite{Konig2014}.

In what  follows we consider  the family  $\hrho_{\theta , \hC}$  generated from
$\hrho_{\theta_0}=\hrho_t$  by  unitaries  of  the  form  $\hU_{\delta\theta}  =
e^{-\frac{\imath}{\hbar} \, \delta\theta \, \hC}$ (with $\delta\theta = \theta -
t$).

\subsection{The rate of change of von Neumann entropy}
\label{sec:vNopensystemsC}

By  exploiting  the  decomposition~\eqref{L1L2L3} with  $\opL_1$,  $\opL_2$  and
$\opL_3$ given in Eqs.~$\eqref{eq:L1L2L3}$, we obtain  that the rate of change of the 
von Neumann entropy $S[\hrho_t] = - \Tr(\hrho_t \ln\hrho_t)$, can be written in  terms of the divergence-based quantum
Fisher information~\eqref{eq:QFIdef_commutator}.  Note first that $\dv*{S[\hrho_t]}{t} =
-\Tr\left(  (   \dv*{\hrho_t}{t}  )   \,  \ln\hrho_t   \right)  =   -  \Tr\left(
\opL[\hrho_t]    \,   \ln\hrho_t    \right)   =    -   \Tr\left(    \hrho_t   \,
\bar{\opL}[\ln\hrho_t] \right)$.  Consequently, from  the expression of $\opL =
\opL_{\mathrm{U}}   +    \opL_{\mathrm{NU}}$   given    by   Eqs.~\eqref{eq:LU} and
\eqref{L1L2L3}-\eqref{eq:L1L2L3},  together  with  the selfadjoint  character  of
$\opL_1$ and $\opL_2$, and the antisymmetry of $\opL_3$, we obtain
\begin{equation}
\label{dSdtgeneral}
\dv{S[\hrho_t]}{t} = \Delta_t - \Psi_t
\end{equation}
with $\Delta_t = -\Tr( \hrho_t \opL_1[\ln  \hrho_t])$ and $\Psi_t = \Tr( \hrho_t
(\opL_2 - \opL_3)[\ln \hrho_t])$. Notice that we used the fact that
\begin{equation}
\label{eq:LUeffetvanishes}
\Tr(\hrho_t \; [\hH , \ln \hrho_t ]) = 0
\end{equation}
(this  can  be   established  by  simple  algebra)  so   that  the  contribution
to~\eqref{dSdtgeneral}   from   the  unitary   evolution,   $\opL_{\mathrm{U}}$,
vanishes.

Now,   from   the   expressions   of   $\opL_1$~\eqref{eq:L1}   and of the DQFI in the form~\eqref{eq:QFIdef_commutator}, we can express the first contribution in the r.h.s.\ of Eq.~\eqref{dSdtgeneral} in terms of the DQFI, as follows 
\begin{equation}
\label{dS1general}
\Delta_t =  \frac12 \sum_{k=1}^K \left. \left(  J_q[\hrho_{\theta , \sqrt{\hbar}
    \hA_k} ; \theta] + J_q[\hrho_{\theta , \sqrt{\hbar} \hB_k} ; \theta] \right)
\right|_{\theta = t},
\end{equation}
where $\hrho_{\theta , \hC}$ are  generated by the unitaries $\hU_{\delta\theta}
= e^{-\frac{\imath}{\hbar}  \, \delta\theta \,  \hC}$ ($\delta\theta =  \theta -
t$), with $\hC = \sqrt{\hbar} \hA_k$  or $\sqrt{\hbar} \hB_k$ being the Hermitian
generators of the unitaries.  From $S[\hrho_{\theta , \hC} \| \hrho_t] = \frac12
\left.  J_q[\hrho_{\theta , \hC} ; \theta] \right|_{\theta = t} \delta\theta^2 +
o(\delta\theta^2)$  (see Appendix~\ref{appA}  or Refs.~\cite{Nagaoka2005, Konig2014}),
we conclude  that $\Delta_t$ is  essentially a measure  of the noise  induced by
unitaries  $\hU_{\delta\theta}$, with  generators  $\sqrt{\hbar} \hA_k$  and
$\sqrt{\hbar}\hB_k$, on  the state  $\hrho_t$.  Accordingly, $\Delta_t$  plays a 
role analogous to  that given  by the  first term, ~\eqref{Flucontribution}, in the r.h.s. of the  classical de  Bruijn
identity.  

In  addition,  let us  emphasize  that  both
quantities $\Delta_t$ and $\Psi_t$ are invariant under the transformations given
by  Eq.~\eqref{Invtransf}.  This  is  due to  the effects~\eqref{Invinfgen}
induced in the superoperators  together with~\eqref{eq:LUeffetvanishes} that
justifies   that  the  energy-conserving 
contributions  coming   from  $\hH^\prime$    induce  no
contribution in $\Psi_t$. 
As a consequence, $\Psi_t$ characterizes the contributions of dissipative forces
to   the    rate   of   change    of   von   Neumann    entropy.    Accordingly,
decomposition~\eqref{dSdtgeneral} reflect  a \  fluctuation-dissipation relation
in the rate of change of the von Neumann entropy for QDSs, analogous to the
one given  by Eq.~\eqref{eq:BruGenR} for  the Shannon entropy rate  in classical
channels corresponding to Ornstein--Uhlenbeck processes.

We highlight that the fluctuation-dissipation relation~\eqref{dSdtgeneral} is
a direct  consequence of the decomposition~\eqref{L1L2L3} for the infinitesimal  generators in the LME.  
 In this respect, we can say  that the decomposition~\eqref{L1L2L3} is  itself a fluctuation-dissipation  relation for QDSs  in its
own,  where  the  infinitesimal  generator   $\opL_1$  is  associated  with  the
fluctuation  part of  the evolution  whereas $\opL_{2+3}$  with the  dissipation
part.

Notice  that  our  way  to  express  the rate  of  change  of  the  von  Neumann
entropy, Eq.~\eqref{dSdtgeneral}, differs from the usual decomposition
\begin{equation}
\label{dSdtPiPhi}
\dv{S[\hrho_t]}{t} =  \Pi_t - \dot\Phi_t
\end{equation} 
given in  terms of the  rate $\Pi_t = \dv*{\Sigma_t}{t}  \geq 0$ of  the entropy
production  $\Sigma_t$,  and the  rate  $\dot\Phi_t  = \dv*{\Phi_t}{t}$  of  the
entropy flux $\Phi_t$~\cite{Landi2020}.  This decomposition  is one way to write
the second law of the thermodynamics.   Although there is a general proposal for
the form of the entropy production $\Sigma_t$ and the entropy flux $\Phi_t$, for
a  general system-environment  evolution (see~\cite{Landi2020} and 
references herein), the entropy production,  in general, depends on the evolved
reduced  state of  the  environment,  which is  not  available  in open  quantum
systems.   However,   for  QDSs  this   problem  was  overcome  long   time  ago
by Spohn in~\cite{Spohn1978},  but  only in  the  cases  when  these semigroups  have  an
invariant state $\hrho^s$,  \ie $\hrho^s = e^{t\opL}\hrho^s, \forall  t \geq 0$.
In  this  situation  the  entropy  production rate  is  $\Pi_t  =  -\dv{S[\hrho_t  \|
    \hrho^s]}{t} \geq 0$.

The decomposition  in~\eqref{dSdtPiPhi} is useful to  study stationary states,
not   necessarily   equilibrium   ones,  arising   when   $\Pi_t   =\dot\Phi_t$.
Furthermore,  these states,  $\hrho^s$, are  equilibrium states  if and  only if
$\Pi_t =  \dot\Phi_t= 0$.   In the  case of  QDSs, $\Pi_t$  is also  a monotonic
convex function  of time as a  consequence of $-S[\hrho_t \|  \hrho^s]$ being an
increasing function of time.  This determines also how the approach to the
stationary state is.

Our decomposition~\eqref{dSdtgeneral},  that it is still valid for  QDSs without a
stationary state, is also useful to study stationarity in this type of
open   systems    but   on   a    different   perspective   from    that   given
by~\eqref{dSdtPiPhi}.   In our  framework,  stationarity  arises when  $\Psi_t$
balances $\Delta_t$, since each DQFI in~\eqref{dS1general} is always positive.
As  we  have  already  established,  our  decomposition~\eqref{dSdtgeneral}  is  a
fluctuation-dissipation relation, therefore the balance between $ \Delta_t $ and
$ \Psi_t $ can be interpreted  as a fluctuation-dissipation equilibrium. We will
confirm this  point of view  in Sec.~\ref{subsec:StatStat} for  quantum Gaussian
channels  that  admit  a  QDS  description~\cite{Heinosaari2010}  that  we  call
Gaussian dynamical  semigroups (GDSs).  This  is the quantum counterpart  of the
fluctuation-dissipation equilibrium balance we found in the classical de Bruijn
equation in Sec.~\ref{subsec:deBruijnB}.

\section{Rate of von Neumann entropy for Gaussian dynamical semigroups}
\label{sec:QdBiGaussian}

\subsection{Gaussian dynamical semigroups}
\label{sec:QdBiGaussianA}

In what follows, we focus on  GDSs, being the attenuator, amplifier and additive
Gaussian noise channels relevant examples~\cite{Heinosaari2010, Giovannetti2010,
  DePalma2016}.  GDSs  are also  useful to describe  damped collective  modes in
deep inelastic collisions~\cite{Sandulescu1987}.

The kinematics of  a quantum Gaussian channel of $n$  bosonic modes is described
by  a $(2  n)$-dimensional vector  of canonically  conjugate operators,  $\hbx =
(\hq_1 , \ldots  , \hq_n , \hp_1 ,  \ldots , \hp_n)^\tp $, such that  $[ \hx_j ,
  \hx_k ]  = \imath  \hbar \JJJ_{jk}  \huno$, with $\JJJ  = \begin{pmatrix}  0 &
  \II_n\\ -\II_n &  0\end{pmatrix}$ the $2 n \times 2  n$ real symplectic matrix
  where $\II_n$ is the $n \times n$ identity matrix,
with  $\JJJ^{-1} =  -\JJJ =  \JJJ^\tp $.   The  dynamics of  a GDS  is given  by
LME~\eqref{eq:mastereq} with a  Hamiltonian up to quadratic order  in $\hbx$ and
linear Lindblad operators,
\begin{equation}
\label{Hquadratic}
\hH =  \frac12 \hbx^\tp   \BB \hbx  + \hbx^\tp  \JJJ  \bxi \quad  \text{and} \quad
\hL_k = \l_k^\tp  \JJJ \hbx,
\end{equation}
respectively, where $\BB \ge  0$ (Hessian matrix), $\bxi$ is an  arbitrary $(2 n)$-dimensional
real  vector,  and  the  $\l_k$s are  $(2  n)$-dimensional  complex  vectors.
Usually    master    equations    of     this    type    are    called    linear
LMEs~\cite{WisemanMilburn2009}.

In  the   Weyl--Wigner  representation,   the  symbol  of  $\hrho_t$ is the  Wigner function $W(\x,t)$,  where $\x =  (q_1 ,
\ldots   ,    q_n   ,   p_1   ,    \ldots   ,p_n)^\tp   $   are    phase   space
coordinates~\cite{Ozorio1998}.   The  time  evolution  of  $W$  is  given  by  a
Fokker--Planck  equation of  the  Ornstein--Uhlenbeck form~\eqref{eq:FPE},  with
drift $\AA$ and diffusion $\DD$ matrices given by
\begin{equation}
\label{AA}
\AA = \JJJ \BB-\CC\JJJ , \qquad \DD = \hbar \Re(\GM),
\end{equation}
$\GM  =  \sum_{k=1}^K  \l_k  \l_k^\dag$  is  the  matrix  that  establishes  the
connection with the  Lindblad operators given Eq.~\eqref{Hquadratic}  and $\CC =
\Im(\GM)$ is the dissipation matrix.  By definition, $\GM \ge 0$ so that
\begin{equation}
\DD - \imath  \hbar\CC \geq 0,
\end{equation}
which   can   be   interpreted   as   a   generalized   fluctuation--dissipation
relation~\cite{WisemanMilburn2009}.   This matrix  inequality implies  that $\DD
\geq 0$, because  $\DD$ and $\CC$ are symmetric and  antisymmetric real matrices
respectively.

A few  observations and  remarks are  in order here.   First, defining  the real
vectors
\begin{equation}
\bSigma_k  =  \sqrt{\hbar}  \Re(\l_k),  \qquad  \bar{\bSigma}_k  =  \sqrt{\hbar}
\Im(\l_k),
\end{equation}
the diffusion matrix can be expressed as 
\begin{equation}
\label{DDdecomposition2}
\DD
= \sum_{k=1}^K   \big(  \bSigma_k
\bSigma_k^\tp  + \bar{\bSigma}_k \bar{\bSigma}_k^\tp  \big),
\end{equation}
If     we      compare     the      expressions     Eqs.~\eqref{DDdecomposition}
and~\eqref{DDdecomposition2},  we  immediately  recognize  in  the  set  $\big\{
\bSigma_k  \,  ,  \,  \bar{\bSigma}_k  \big\}_{k=1,\ldots,K}$  Langevin  forces.
However, all these  forces are not necessarily linear independent  like those in
Eq.~\eqref{DDdecomposition}.   Because the  complex vectors  $\l_k$ are  usually
linearly independents (with $K  \le 2 n$), $K$ is the rank  of the matrix $\GM$.
But the rank of the matrix $\DD$ expressed in Eq.~\eqref{DDdecomposition2} could
range from $K  \leq \mbox{rank}(\DD) = M \leq  \min\{ 2n , 2K \}$  being $K$ for
example when  all pairs $\bSigma_k,  \bar{\bSigma}_k$ are composed  with vectors
proportional to  each other, and  being $M=2K \le 2n$  for example when  all these
vectors  are linearly  independent.  
Conversely, the  rank  of the  dissipation
matrix $\CC = \Im(\GM)$ could range  from $0 \leq \mbox{rank}(\CC)\leq \min\{ 2n
, 2K  \}$.  Therefore, it  is not  possible in GDSs  to have a  dynamics without
diffusion while it is  possible to have a dynamics without  dissipation as it is
the case of a  quantum diffusion process.  The latter is a  particular case of a
GDS  where $\DD  =  \frac12 \II$  and  $\CC =  0$, which  is  precisely the  GDS
considered in context of quantum de Bruijn identity in~\cite{Konig2014}.

Remind  that in  GDSs  the  Fokker--Plank equation  Eq.~\eqref{eq:FPE}
propagates  the  Wigner  function  $W(\x,t)$, that  it  is  a  quasi-probability
distribution that  describes completely  the quantum  state of  the system.
However,  in the  classical mechanics  context the  same Fokker--Plank  equation
Eq.~\eqref{eq:FPE}, with the drift $\AA$ and diffusion $\DD$ matrices given
in~\eqref{AA}, is  also used, for  example, to  describe the Brownian  motion in
a harmonic  potential. In  this case  the drift  force in
phase space can be splitted into $\vF_{dr}(\x) = \vF_S(\x) + \vF_{AS}(\x)$,
where $\vF_{AS}(\x)  = \JJJ  \BB \x-\bxi$  and $\vF_S(\x) =  - \CC  \JJJ \x$  are the
conservative  and dissipative  forces, respectively.   Note,  that in  this case,  as
commented     below   Eq.~\eqref{trAinterpretedFlux},
$\pdv*{\vF_{AS}(\x)}{\x^\tp} =  \tr(\AA_{AS}) = \tr(\JJ \BB)
=   0$  because   $\JJJ$  is   anti-symmetric  and   $\BB$  is   symmetric,  and
$\pdv*{\vF_S(\x)}{\x^\tp} = \tr(\AA_S) = - \tr(\CC \JJ)$.

Finally, using successively  (i) the expression for the drift  matrix $\AA =
  \JJJ \BB - \CC \JJJ$ in~\eqref{AA} so that $- \pdv*{(\AA \x - \bxi)}{\x^\tp} =
  - \tr(\AA) = \tr( \CC \JJJ)$, (ii) the quadratic classical Hamiltonian $H(\x)=
  \frac12 \x^\tp  \BB \x +  \x^\tp \JJJ \bxi$ --the  Weyl symbol of  the quantum
  Hamiltonian  Eq.~\eqref{Hquadratic}--  together with  $\JJJ^{-1}  =  - \JJJ  =
  \JJJ^\tp$ so that $ (\AA \x - \bxi)^\tp = - \pdv*{H(\x)}{\x^\tp} \JJJ - \left(
  \CC \JJJ \x \right)^\tp$  and (judiciously) $a^\tp b = b^\tp a  = \tr( b a^\tp
  )$ for any vector,
the Fokker--Planck equation in~\eqref{eq:FPE} for the Wigner function 
$W(\x,t)$ of the system can be rewritten under the form
 \begin{align}
 \label{FPEGDS}
 \dv{W({\bf   x},t)}{t}&  =   [H(\x),W(\x,t)]_{cl}+\frac12  \tr(\DD   \HH[W({\bf
     x},t)])\nonumber\\
 & + \tr(\CC \JJ) W(\x,t) + \tr\left( \CC \JJ \; \x \pdv{W(\x,t)}{\x^\tp}
 \right),
 \end{align}
 where the first term is  the Poisson bracket, $\tr\left( \JJJ \pdv{W(\x,t)}{\x}
 \frac{\partial H}{\partial  \x^T}\right)$, between the Wigner  function and the
 quadratic classical Hamiltonian.

Let  us  now  write  the  corresponding  LME of  a  GDS  in  terms  of  the
decomposition  of  the  infinitesimal generator  given  Eq.~\eqref{L1L2L3}.
More precisely, in Appendix~\ref{appB}, we show that
\begin{subequations}
\label{eq:L1L2L3GDS}
\begin{align}
\label{eq:L1GDS}
\opL_1[\hO] &=   \frac12 \tr(\DD \hHH[\hO]), \\
\label{eq:L2GDS}
\opL_2[\hO] &= \frac12  \tr( \JJJ \CC) \hO, \, \text{and} \\ 
\label{eq:L3GDS}
\opL_3[\hO] &= \tr(\CC \JJJ \hMM[\hO]) +  \opL_2[\hO], 
\end{align}
\end{subequations} 
where we define the superoperator matrices
\begin{subequations}
\label{defhatHM}
\begin{eqnarray}
\label{defhatH}
\hHH  [\hO] &=& -\frac{1}{\hbar^2} [\JJJ  \hbx , [(\JJJ  \hbx)^\tp   ,  \hO]]=
\pdv{}{\hbx}\pdv{}{\hbx^\tp},\;\;\mbox{and}\\  
\hMM[\hO]  &=&  \frac{\imath}{\hbar}  \hbx [(\JJJ  \hbx)^\tp  ,  \hO]=\hbx\,\pdv{}{\hbx^\tp},
\label{defhatM}
\end{eqnarray}
\end{subequations} 
with the  matrix notation:  $([\hbx , [\hby^\tp  , \hO]])_{ij} =  [x_i ,  [y_j ,
    \hO]]$, and $(\hbx \,  [\hby^\tp , \hO])_{ij} = x_i [y_j  , \hO]$.  The last equalities in 
    Eqs.\eqref{defhatHM} follows from the identity $(\imath/\hbar)[(\JJJ\hbx)^\tp,\hat O(\hbx)]=
    \partial \hat O(\hbx)/\partial \hbx^\tp$ given in~\cite{Transtrum2005}. Then, the
linear LME is
\begin{align}
	\label{linearLME}
	\dv{\hrho_t}{t}&=\frac{1}{\imath\hbar}[\hat H,\hrho_t]+
	\frac{1}{2}\tr\left(\DD\hHH[\hrho_t]\right)\nonumber\\
	&+\tr(\CC\JJJ)\hrho_t+\tr(\CC \JJJ \hMM[\hrho_t]).
\end{align} 
For   one   mode  ($n=1$),   this   expression   reduces  to   that   given
in~\cite{Sandulescu1987}.

Notice  that~\eqref{FPEGDS}   is   nothing   but   the   Weyl-Wigner
representation of~\eqref{linearLME}.   Moreover,  a direct
comparison  of both  evolution  equations 
  follows from the correspondences
\begin{subequations}
\label{rules}
\begin{eqnarray}
\hrho_t & \: \leftrightarrow \: & W(\x,t),
\label{rule1}\\
\hbx & \: \leftrightarrow \: &\x, 
\label{rule2}\\
(\imath/\hbar)[(\JJJ\hat{\bf  x})^\tp\,,\cdot\,]=\partial/\partial{\hbx^\tp}&  \:
\leftrightarrow \: & \pdv*{\x^\tp}\\ (1/\imath  \hbar)[ \cdot \, , \,
  \cdot ]& \: \leftrightarrow \: & [\cdot  \, , \, \cdot ]_{cl},
\label{rule3}\\
\Tr(\cdot) & \: \leftrightarrow \: &\int_{\Rset^N} \cdot \: \dd^{2n} \x.
\label{rule4} 
\end{eqnarray}
\end{subequations}
In this  way, we  can go  back and  forth between  both equations~\eqref{FPEGDS}
and~\eqref{linearLME}.  
In the next section, we will see that these correspondences also allows to
go   back   and   forth   between    the   generalized   classical   de   Bruijn
identity~\eqref{eq:BruGen}  and  the quantum  de  Bruijn  identity for  the  von
Neumann entropy rate in GDSs Eq.~\eqref{rateSfinal}.

\subsection{Quantum de Bruijn identity for Gaussian dynamical semigroups}
\label{Sec:QuandBiGDS}

As we have seen, decomposition~\eqref{L1L2L3} for a GDS splits the dynamics into
diffusive and dissipative contributions, given by the corresponding superorators
$\opL_1$~\eqref{eq:L1GDS},    and    $\opL_2$     and    $\opL_3$    given    in
Eqs.\eqref{eq:L2GDS}    and~\eqref{eq:L3GDS},     respectively.     From    this
decomposition,  we obtain  the  quantum  version of  the  generalized de  Bruijn
identity for GDSs,
\begin{equation}
\label{rateSfinal}
\dv{S[\hrho_t]}{t}  =  \Delta_t-\Psi_t=  \frac12  \tr( \DD  \JJ_q[\hrho_t]  )  -
\tr(\CC\JJJ \MM[\hrho_t] ),
\end{equation}
where 
\begin{equation}
\label{quaFishermat}
\JJ_q[\hrho_t] =  - \Tr( \hrho_t \hHH[\ln \hrho_t])
\end{equation} 
is   the  DQFI   matrix,   a   quantum  version   of   the  Fisher   information
matrix~\eqref{claFishermat}, and
\begin{equation}
\label{defMsuop}
\MM[\hrho_t] =  - \Tr(  \hrho_t \hMM[\ln  \hrho_t] ).
\end{equation}  
We observe  that the  result~\cite[Eq.(82)]{Konig2014} obtained for  the quantum
diffusion  process (for  which $\DD  =  \frac12 \mathbb{1}$  and $\CC  = 0$)  is
recovered from~\eqref{rateSfinal} evaluated at $t  = 0$ and performing the trace
operation.    This  is   the  quantum   analog  of   the  classical   de  Bruijn
identity~\eqref{eq:dBi}.  As with its classical counterpart, this result is used
in  the  derivation  of   quantum  entropy  power  inequalities~\cite{Konig2013,
  DePalma2014}.

The  diffusion   contribution  $\Delta_t$  in~\eqref{rateSfinal}  can   also  be
expressed   as~\eqref{dS1general}    with   $\hA_k    =   \frac{1}{\sqrt{\hbar}}
\bSigma_k^\tp \JJJ \hbx$ and $\hB_k = \frac{1}{\sqrt{\hbar}} \bar{\bSigma}_k^\tp
\JJJ \hbx$,  where $\bSigma_k  = \Re(\l_k)$  and $\bar{\bSigma}_k  = \Im(\l_k)$.
Therefore, the family of density operators in~\eqref{dS1general}, \ie
\begin{equation}
\label{hrhothetaC}
\hrho_{\theta,  \hC} =  \hU_{\delta\theta} \, \hrho_t  \, \hU_{\delta\theta}^\dag =
\hT_{\bseta} \, \hrho_t \, \hT^\dag_{\bseta},
\end{equation}
with $\hC  = \bSigma_k^\tp  \JJJ \hbx$ or  $\bar{\bSigma}_k^\tp \JJJ  \hbx$, are
generated    by    phase    space    translation    operators~\cite{Ozorio1998},
$\hT_{\bseta}$,  with  $\bseta  = \bSigma_k  \delta\theta$  or  $\bar{\bSigma}_k
\delta\theta$.   These translations  in  the phase  space  can be  directly
related to the Langevin forces  $\O_k = \bSigma_k, \bar{\bSigma}_k$.  Indeed, if
we consider the  classical counterpart of the  Hamiltonian generator $\hC$,
\ie  $C =  \O_k^\tp \JJJ  \x$,  from the  Hamilton  equation of  motion we  have
$\dv{\x}{t}  =  \JJJ \frac{\partial  C}{\partial{\x}}  =  \O_k$.  Following  the
discussion below~\eqref{dS1general}, we conclude that in this case the diffusion
term $\Delta_t$  in~\eqref{rateSfinal}, is  a measure of  the noise  produced by
these  Langevin   forces,  $\bSigma_k   =  \Re(\l_k)$  and   $\bar{\bSigma}_k  =
\Im(\l_k)$, in the quantum evolution.

Finally,   from   the   correspondences~\eqref{rules}  and replacing  the von  Neumann entropy  by the  Shannon
entropy into the  quantum de Bruijn identity~\eqref{rateSfinal},  we can recover
the classical  de Bruijn equation~\eqref{eq:BruGen}, and  conversely  we   can
obtain~\eqref{rateSfinal}  from~\eqref{eq:BruGen}.            Moreover,
the correspondences in~\eqref{rules}  imply
$\JJ_q[\hrho_t] \:  \leftrightarrow \: \JJ[p_t]$  for the quantum  and classical
information       matrices        given       in       Eqs.~\eqref{quaFishermat}
and~\eqref{claFishermat}, respectively, and  $\MM[\hrho_t] \: \leftrightarrow \:
- \int_{\Rset^N} W(\x,t)  \; \x  \; \pdv{\ln W(\x,t)}{\x^\tp}  \; \dd^{2n}  \x =
\II$ (see Eq.~\eqref{identity}), as well.   Therefore, we conclude that the
dissipation term $\Psi_t$ in~\eqref{rateSfinal} measure
 the total  amount of  change of  the density  operator due  to the
dissipative force $\hat\vF_S(\x) = - \CC \JJJ \hbx$.

\subsection{Gaussian states}
\label{Sec:QuandBiGDSGau}

The  Fokker--Planck   equation~\eqref{eq:FPE}  associated  with  a   linear  LME
propagates any initial Wigner function  that describes the initial quantum state
of  the   system~\cite{Nicacio2010}.   Generally,  these   are  quasiprobability
distributions,   \ie   they   can    take   negative   values   but   still
$\int_{\Rset^N}\,W(\x,t)     \;    \dd^{2n}     \x    =     1$.     Here,     we
discuss~\eqref{rateSfinal} specialized to Gaussian  states $\hrhog_t$, which are
quantum states  whose Wigner  functions are probability  distributions.  Indeed,
the Wigner  function of a Gaussian  state $\hrhog_t$ is a  multivariate Gaussian
distribution with mean $\expval{\hbx}_t  = \Tr(\hrho_t \hbx)$ and covariance
  matrix $\VVt  = \frac{1}{2\hbar}  \Tr( \hrho_t(\hbx -  \expval{\hbx}_t)(\hbx -
  \expval{\hbx}_t)^\tp )$,
\begin{equation}
\label{eq:WignerGauss}
W_G(\x,t)  =  e^{-\frac{1}{2\hbar}(\x  - \expval{\hbx}_t)^\tp  \VVt^{-1}  (\x  -
  \expval{\hbx}_t)} / \W_t,
\end{equation}
where  $\W_t =  (2  \pi  \hbar)^n \sqrt{\det  (\VVt)}$.   Moreover, the  density
operator of a Gaussian states can be expressed as~\cite{Holevo2019, Banchi2015}
\begin{equation}
\label{denopGS}
\hrhog_t  =  {e^{-\frac{1}{2\hbar}(\hbx  -  \expval{\hbx}_t)^\tp  \UUt  (\hbx  -
    \expval{\hbx}_t)}} / \Z_t,
\end{equation}
with $\Z_t = \sqrt{\det\left( \VVt  + \frac{\imath}{2} \JJJ \right)}$ and $\UU_t
= 2  \imath \JJJ  \coth^{-1}\left( 2  \imath \VV_t  \JJJ \right)$.   Notice that
matrices $\VVt$ and $\UUt$ satisfy
\begin{equation}
\label{UVJJVU}
\UUt \VVt \JJJ = \JJJ \VVt \UUt.
\end{equation}   
The main reason to focus on Gaussian  states is that they represent the universe
of stationary states in GDSs~\cite{Carmichael1999}.

The evolution of the covariance matrix and  first moments of any state in linear
LME is then determined by the differential equations
\begin{equation}
\label{timeLyapunov}
\frac12 \dv{\VVt}{t} =  \frac{\DD}{2 \hbar} + \frac12 \left( \VVt  \AA^\tp + \AA,
\VVt \right),
\end{equation}
and
\begin{equation}
\label{timeMean}
\dv{\expval{\hbx}_t}{t}  = \AA \expval{\hbx}_t -  \bxi,
\end{equation}
respectively, with $\AA$ and $\DD$ given by Eq.~\eqref{AA} and $\bxi$ as in
Eq.~\eqref{Hquadratic}.  These equations completely determine the evolution
of Gaussian states in GDSs.

The Shannon entropy of a Gaussian  state $\hrhog_t$ is well defined (since $W_G$
is a probability distribution), and given by $h[\mbox{$\scriptstyle{W_G}$}]
= \frac12  \ln\det(\VVt) + n \ln(2  \pi e)$.  Therefore, the  change of
  rate of the Shannon entropy
is given by
\begin{equation}
\label{dhWGdt}
\dv{h[\mbox{$\scriptstyle{W_G}$}]}{t}      =      \frac12     \tr\left(      \DD
\frac{\VVt^{-1}}{\hbar} \right) - \tr(\JJJ \CC),
\end{equation}
which      is      directly       obtained      from~\eqref{eq:BruGen}      with
$\JJ[\mbox{$\scriptstyle{W_G}$}]  = \frac{\VVt^{-1}}{\hbar}$  and $\tr(\AA)  = -
\tr(\JJJ\CC)$.    As   in~\eqref{rateSfinal},    there   is    no   contribution
in~\eqref{dhWGdt} of the Hamiltonian evolution.

We   emphasize here  that~\eqref{dhWGdt}    is    simply    the   classical    Bruijn
identity~\eqref{eq:BruGen} for  a probability  distribution that  corresponds to
the Wigner function of a Gaussian quantum state.

In  Sec.~\ref{subsec:deBruijnB}  we  showed  that the  first  term  in  the
classical  de   Bruijn  identity   in~\eqref{eq:BruGen}  can  be   rewritten  as
in~\eqref{Flucontribution}.  However, in  this case  it is  more interesting  to
rewrite  the  first  term  in~\eqref{dhWGdt}  using  the  expression  for  $\DD$
in~\eqref{DDdecomposition2} that contains the Langevin forces $\O_k = \bSigma_k,
\bar{\bSigma}_k$, \ie
\begin{equation}
\frac12  \tr\left(  \DD  \frac{\VVt^{-1}}{\hbar} \right)=  \frac12  \sum_{k=1}^K
\left(   J[W^{(\theta   ,   \bSigma_k)}_G  ;   \theta]   +\left.J[W^{(\theta   ,
    \bar{\bSigma}_k)}_G ; \theta] \right) \right|_{\theta=t}.
\end{equation}
Here,  $\O_k^\tp \JJ[W_{G}]  \O_k =  \left.  J[W^{(\theta  , \O_k)}_G  ; \theta]
\right|_{\theta = t}$, where $J[W_G ;  \theta] = \int_{\Rset^{2n}} W_G \, \left(
\pdv*{\ln W_G}{\theta} \right)^2 \dd^{2n} \x$,  is the Fisher information of the
Wigner function  of the  translated state given  by Eq.~\eqref{hrhothetaC},
$W_G^{(\theta , \O_k)} = W_G(\x - \delta\theta \O_k , t)$.

Finally,  for   Gaussian  states,   we  obtain  that   the  quantum   de  Bruijn
identity~\eqref{rateSfinal} reduces to
\begin{equation}
\label{dSdtforGSa1}
\dv{S[\hrhog_t]}{t}  =  \frac12  \tr\left(   \DD  \frac{\UUt}{\hbar}  \right)  -
\tr\left( \JJJ \CC \UUt \VVt \right),
\end{equation}
where we used that  (i) $\hHH[\ln \hrhog_t ] = -  \frac{1}{\hbar} \UUt \; \huno$
and  $\hMM[\ln  \hrhog_t]  = \frac{1}{\hbar}  \hbx(\hbx  -  \expval{\hbx}_t)^\tp
\UUt$, so that  $\MM[\hrhog_t]=-\VVt\UUt-\frac{\imath}{2}\JJJ\UUt$, and that, in
general, $\Tr(\hrho_t  \hbx \hbx^\tp  ) - \expval{\hbx}_t  \expval{\hbx}_t^\tp =
\VVt + \frac{\imath}{2} \JJJ$, and that (ii) $\tr(\UUt\JJJ) = 0$, because $\UUt$
is     symmetric    and     $\JJJ$     antisymmetric,    and (iii) relation~\eqref{UVJJVU}.

\subsection{Stationary states in Gaussian dynamical semigroups}
\label{subsec:StatStat}

In GDSs if  there  exists a  stationary state,
  $\hrhog^{\scriptstyle  S}$, this  one is  unique  and for  any initial  state
$\hrho_0$,   we  have   $\hrho_t\underset{t  \rightarrow   +\infty}{\rightarrow}
\hrhog^{\scriptstyle S}$~\cite{Frigerio1977,Frigerio1978,Carmichael1999}.  In particular, starting with an
initial Gaussian  state, the  evolved states remain Gaussian. Therefore, 
stationary states necessarily are Gaussian.   Remarkably, the
quantum de Bruijn identity  given by expression~\eqref{rateSfinal}, for
  whatever  initial   state,   together  with   the  one   given by
  Eq.~\eqref{dSdtforGSa1}  for an  initial  Gaussian  state, allow  a
description of stationary situations.  Indeed, from our approach we observe that
the stationary state arises when the  environment, described by $\DD$ and $\CC$,
allow the  balance between the  diffusion term $\Delta_t  > 0$
and  the dissipative term $\Psi_t$ \ie
$\dv{S[\hrho_t]}{t}\underset{\scriptstyle  t  \rightarrow  +\infty}{\rightarrow}
0$.  While  $\Delta_t$ describes the increase  in noise due to  Langevin forces,
$\Psi_t$   represents  the   contribution   due  to   the  dissipative   forces.
Particularly,    for   any    initial    Gaussian    state,   the    convergence
$\hrhog_t\underset{t \rightarrow  +\infty}{\rightarrow} \hrhog^{\scriptstyle S}$
happens     in      such     a      way     that      $\dv{S[\hrhog_t]}{t}     -
\dv{h[\mbox{$\scriptstyle{W_G}$}]}{t}\underset{t\rightarrow
  +\infty}{\rightarrow} 0$.  This is a consequence of
\begin{equation}
\label{dSdtforGSa2}
\dv{S[\hrhog_t]}{t} = \dv{h[\mbox{$\scriptstyle{W_G}$}]}{t}  + \frac12 \tr\left(
\Thet_t \dv{\VVt}{t} \right),
\end{equation}
and     $\dv{\VV_t}{t}\underset{t    \rightarrow     +\infty}{\rightarrow}    0$
in~\eqref{timeLyapunov}       when        $\hrhog_t\underset{t       \rightarrow
  +\infty}{\rightarrow}          \hrhog^{\scriptstyle         S}$.           The
relation~\eqref{dSdtforGSa2}  between  von  Neumann and  Shannon  entropy  rates
results from
\begin{subequations}
\label{dSdtforGSa3}
\begin{align}
\dv{S[\hrhog_t]}{t}   &=  \frac12   \tr(   \UUt  \dv{\VVt}{t}   )  , 
\label{dSdtforGSa3a}\\
\dv{h[\mbox{$\scriptstyle{W_G}$}]}{t}     &=    \frac12     \tr(    \VVt^{-1}
\dv{\VVt}{t}),
\label{dSdtforGSa3b}
\end{align}
\end{subequations}
and   the  expansion   $\UUt  =   \VV^{-1}_t  +   \Thet_t$,  where   $\Thet_t  =
\sum_{m=1}^{+\infty}  \frac{2   \imath  \JJJ}{2m+1}  \left(   \frac{\imath  \JJJ
  \VV^{-1}_t}{2}      \right)^{2m+1}$.       To      recover      de      Bruijn
identity~\eqref{dSdtforGSa1} from the time derivative of $S[\hrhog_t]$ given
  Eq.~\eqref{dSdtforGSa3a}, we  use~\eqref{timeLyapunov},  $\tr(
\UUt \AA \VVt) = \tr(\UUt \VVt \AA^\tp )$, $\tr\left( \JJJ \CC \UUt \VVt \right)
= \tr\left( \CC \JJJ \VVt \UUt \right)$ and $\tr( \JJJ \BB \VVt \UUt) = 0$.  The
time   derivative  of   $h[\mbox{$\scriptstyle{W_G}$}]$  in~\eqref{dSdtforGSa3b}
directly     follows    from     $\dv{(\ln\det(\VVt))}{t}    =     \tr(\VVt^{-1}
\dv{\VVt}{t})$~\cite[Chap.~1 \&~9]{MagnusNeudecker1999}.

The covariance matrix $\VV^S$ of a stationary state $\hrhog^{\scriptstyle S}$ is
a solution of a  Lyapunov equation~\cite{Nicacio2016}, \ie setting $\dv{\VVt}{t}
=  0$   in~\eqref{timeLyapunov} so,
\begin{equation}
\label{Lyapunoveq}
\frac{\DD}{\hbar} + \VV^S (- \BB \JJJ - \JJJ \CC) + (\JJJ \BB - \CC \JJJ ) \VV^S
= 0.
\end{equation}
The solution of this Lyapunov equation gives $\VV^S$ as a function of $\BB$,
$\DD$ and $\CC$,  and this dependency determines the  type of Gaussian state
 $\hrhog^{\scriptstyle  S}$ is.  Notice that,  according to~\eqref{dhWGdt},
$\dv{h[\mbox{$\scriptstyle{W_G}$}]}{t} = 0$  is equivalent to the  trace of this
Lyapunov  equation, since  $\tr(\BB\JJJ)=0$.   Therefore, there  is a  direct
  connection between  the Lyapunov  equation that  determines $\VV^S$,  and hence
  $\hrhog^{\scriptstyle S}$, and the stationarity of the Shannon entropy for the
  Wigner     function    of     $\hrhog^{\scriptstyle     S}$,    {\it     i.e.}
  $\dv{h[\mbox{$\scriptstyle{W_G}$}]}{t} = 0$.  Besides, for a stationary state
we    have    $\Delta_t     =    \Psi_t$    in~\eqref{dSdtforGSa1},    therefore
$\dv{S[\hrhog_t]}{t} = 0$. However, let us see  that, for a family of thermal
  states, there is another way to  establish the stationarity of the von-Neumann
  entropy, {\it i.e.}  $\dv{S[\hrhog_t]}{t} = 0$, which has  a direct connection
  to the  Lyapunov equation for $\VV^S$.  Indeed, rewriting~\eqref{dSdtforGSa1}
as
\begin{equation}
\dv{S[\hrhog_t]}{t}   =   \tr(  \UUt   \VVt
\left(\frac{1}{2\hbar}  \VVt^{-1} \DD  -  \JJJ \CC  \right)  ),
\end{equation}
we  obtain  a solution  of
$\dv{S[\hrhog_t]}{t}  =  0$  solving  the  Lyapunov  equation  
\begin{equation}
\frac{1}{2\hbar}
(\VV^S)^{-1}  \DD -  \JJJ  \CC  = 0.
\label{Lyaeqther}
\end{equation}
Let us see that this equation is a particular instance of Eq.~\eqref{Lyapunoveq} whose solution corresponds to 
the covariance matrix, as a function of $\DD$ and $\CC$, of a family of thermal states.
Indeed, the solution  is given by $\VV^S  = - \frac{1}{2\hbar} \DD  \CC^{-1} \JJJ$, with
the condition $\DD \JJJ  \CC = \CC \JJJ \DD$ for $\VV^S$  to be symmetric. 
This
covariance matrix  $\VV^S = \VV^{th}$ is  indeed that of the  thermal state
$\hrhog^{th} = \frac{e^{-\beta \hH}}{\Z^{th}}$,  where $\Z^{th} = \Tr( e^{-\beta
  \hH})$, with $\beta$ the  inverse temperature, and $\hH$
given by~\eqref{Hquadratic}  where $\bxi$  must be  zero~\footnote{In GDS  the time
dependence    of   first    moments    is   $\expval{\hbx}_t    =   e^{\AA    \,
  t}(\expval{\hbx}_{\hrho_0} - \AA^{-1} \bxi) + \AA^{-1} \bxi$.  Because the GDS
has  a  stationary   state,  $\AA$  is  asymptotically   stable,  then  $\lim_{t
  \rightarrow +\infty} \expval{\hbx}_t = \AA^{-1} \bxi = - \bxi$, where the last
equality    results   for    a    thermal   state    constructed   with    $\hH$
in~\eqref{Hquadratic}.  But $\AA^{-1}\bxi = -\bxi$ implies  $\BB = - \JJJ \CC \JJJ
+ \JJJ$ that is not a symmetric matrix. Then, we must have $\bxi = 0$}. This is
because  the  Lyapunov   equation~\eqref{Lyapunoveq} for  this  thermal   state
 reduces  to~\eqref{Lyaeqther}, using
that  $\UU^{th}/\hbar =  \beta \BB$  (with $\UU^{th}$ the  matrix in~\eqref{denopGS})
together with $  \UU^{th} \VV^{th} \JJJ = \JJJ \VV^{th}  \UU^{th}$ and $\DD \JJJ
\CC = \CC \JJJ \DD$.  An example  of equilibrium state of this type is that
appearing in  the quantum optical master  equation, for which $\DD  = \gamma\II$
and  $\JJ \CC  = \alpha  \II$, where  $\gamma  \geq 0$  and $\alpha$  is a  real
number~\cite{GardinerZoller2004}.

\section{Conclusions}
\label{sec:conclusions}

For  memoryless  quantum  channels,  whose  dynamics  is  described  by  quantum
dynamical semigroups,  we have provided a  decomposition given in~\eqref{L1L2L3}
of the infinitesimal  generator of the non-unitary evolution.   This extends the
concepts of diffusion and dissipation in  this kind of open quantum systems and
allows us  to obtain a compact  formula for the  rate of change of  the von
Neumann  entropy  of the  quantum  states~\eqref{dSdtgeneral}.  
We emphasize that this constitutes one of the main results of our contribution. The first  term
$\Delta_t$  of that  formula clarifies  that $\opL_1$  in~\eqref{L1L2L3} is  the
infinitesimal  generator of  diffusion  in  the dynamics,  and  the second  term
$\Psi_t$  indicates   that  $\opL_{2+3}$  is  the   infinitesimal  generator  of
dissipation.  Indeed,  while the positive contribution  $\Delta_t$, rewritten as
in~\eqref{dS1general},  corresponds to  the increase  of noise  measured by  the
divergence-based  quantum Fisher  information,  the term  $\Psi_t$ measures  the
contribution of dissipative forces to the rate of change of von Neumann entropy.

We have then  focused on channels that form Gaussian  dynamical semigroups, paramount in the description of the most useful quantum channels in the data-transmission and data-processing systems in modern quantum information theory. For
these  channels, we  have first  obtained  the expression  of the  infinitesimal
generators of  the dynamics~\eqref{eq:L1L2L3GDS}  from our  decomposition, 
where the dependence with the diffusion and dissipative matrices appears explicitly.   
Besides, from  this  decomposition,  we have  provided  a series  of
correspondences~\eqref{rules} that  maps the Lindblad master  equation for the
evolution  of the  density  operator~\eqref{linearLME}  into the  Fokker--Planck
equation  for the  evolution  of the  Wigner  function~\eqref{FPEGDS}, and  vice
versa.  Finally, we obtained the rate  of change of the von Neumann entropy
for   Gaussian   dynamical   semigroups~\eqref{rateSfinal},  where   the   first
term of our identity quantifies  the contributions of diffusion due
to  Langevin  forces,  whereas  the  second  one  is  the  contribution  due  to
dissipative forces.  Identity~\eqref{rateSfinal} is  nothing but the quantum
  counterpart of the generalized classical de Bruijn identity~\eqref{eq:BruGen},
  which   can  also   be  obtained   from  the   latter  from   the  series   of
correspondences~\eqref{rules} (and vice versa).

In  addition, we  have provided  a novel  perspective to  study stationarity  of
quantum dynamical semigroups and in particular for
Gaussian dynamical  semigroups. Such a  study is usually addressed  from the
  balance of  the rates of  the entropy production  and entropy flux,  whereas we
  considered here  the balance between  diffusion and dissipation  in such
quantum channels.

Finally, we  highlight that~\eqref{dSdtgeneral} (with $\Delta_t$ given by Eq.~\eqref{dS1general}) is a  
quantum  de  Bruijn  identity  valid  for  any  quantum
dynamical semigroup. It is remarkable that this identity has a very similar structure to that of the 
classical generalized de Bruijn  identity in~\eqref{eq:BruGenR}, even in the case of finite dimensional quantum systems.  
Thus, we believe that, like the quantum de Bruijn identity in Eq.~\eqref{rateSfinal}, 
without dissipation, is a  key  ingredient  in   the  derivation  of  entropy  power
inequalities in  continuous variable systems, Eq.~\eqref{dSdtgeneral}  with null
dissipative term, {\it i.e.} $\Psi_t=0$, could be useful to  define and prove entropy power inequalities
in finite-dimensional systems.  
Besides, the quantum de Bruijn identity~\eqref{dSdtgeneral} reveals the central role played by the divergence-based 
quantum Fisher information in the quantification of the amount of noise produced by the random effects 
into the dynamics suffered by systems that implement quantum communication channels.
In this respect, our framework opens a new avenue in the study of time evolution in quantum channels from a fluctuation-dissipation perspective that allows to quantify the degradation induced by noise in the information transmitted.

\begin{acknowledgments}
FT  acknowledges financial  support from  the Brazilian  agencies FAPERJ,  CNPq,
CAPES and the  INCT-Informa\c{c}\~ao Qu\^antica.  GMB is  partially supported by
projects ``Indagine sulla struttura logica  e geometrica soggiacente alla teoria
dell'informazione quantistica''  funded by  Ministero dell'Universit\`a  e della
Ricerca  (Italy) and  PICT2018-1774 funded  by Agencia  Nacional de  Promoci\'on
Cient\'ifica y  Tecnol\'ogica (Argentina).  SZ  has been partially  supported by
the  LabEx  PERSYVAL-Lab  (ANR-11-LABX-0025-01)  funded by  the  French  program
Investissement  d\^aavenir.  MP  and  GMB  are members  of  the research  projects
PIP-0519 from CONICET (Argentina) and 11/X812 from UNLP (Argentina).
\end{acknowledgments}

\hfill


\appendix

\section{Proof of Eq.~\eqref{eq:QFIdef_commutator}}
\label{appA}

Let $\hU_{\delta\theta}$ be  a unitary operator describing the  evolution of the
state of a system with respect to a parameter $\delta\theta= \theta - \theta_0$,
\ie    $\hrho_{\theta}    =    \hU_{\delta\theta}   \,    \hrho_{\theta_0}    \,
\hU_{\delta\theta}^\dag$ with  $\hU_{\delta\theta} \,  \hU_{\delta\theta}^\dag =
\hU_{\delta\theta}^\dag  \, \hU_{\delta\theta}  =  \huno$ the  unit operator,  $
\hU_0 = \hat{1}$, and $\theta_0$ a fixed value.

Let us  consider the quantum relative  entropy of the state  $\hrho_\theta$ from
$\hrho_{\theta_0}$ that serves as a reference,
\begin{equation}
\label{eq:QRE}
S\left[  \hrho_\theta  \|  \hrho_{\theta_0}  \right]  =  \Tr\left(  \hrho_\theta
\left(\ln \hrho_\theta - \ln \hrho_{\theta_0} \right) \right).
\end{equation}
Notice  that the  quantum relative  entropy is  positive and  zero, its  minimum
w.r.t, $\theta$, if  and only if $\hrho_\theta =  \hrho_{\theta_0}$.  Thus, both
the  relative entropy  and its  first  derivative w.r.t.,  $\theta$ vanishes  at
$\theta = \theta_0$.

In  order  to obtain  the  expression~\eqref{eq:QFIdef_commutator},  we have  to
calculate   the  second   derivative  of   the  quantum   relative  entropy   at
$\theta=\theta_0$.

The first derivative of the relative entropy is given by
\begin{eqnarray}
\label{eq:firstderivative}
\dv{\theta}  S\left[ \hrho_\theta  \| \hrho_{\theta_0}  \right]  &=& \Tr\left(
\dv{\hrho_\theta}{\theta}  \left( \ln  \hrho_\theta -  \ln\hrho_{\theta_0}
\right)  \right)\nonumber\\
&& +  \Tr\left( \hrho_\theta  \dv{\theta}  \ln \hrho_\theta
\right).
\end{eqnarray}
Because $\hU_{\delta\theta}$ is unitary,  $\ln \hrho_\theta = \hU_{\delta\theta}
\, \ln \hrho_{\theta_0} \, \hU_{\delta\theta}^\dag$.  Thus,
\begin{equation}
\label{dlnrhodtheta}
\dv{\theta}  \ln   \hrho_\theta  =   \dv{\hU_{\delta  \theta}}{\theta}   \,  \ln
\hrho_{\theta_0}   \,  \hU_{\delta\theta}^\dag   +  \hU_{\delta\theta}   \,  \ln
\hrho_{\theta_0} \, \dv{\hU_{\delta\theta}^\dag}{\theta}.
\end{equation}

Now,   tracing   Eq.~\ref{dlnrhodtheta}   and  using   successively   (i)   both
$\hrho_\theta = \hU_{\delta\theta} \hrho_{\theta_0} \hU_{\delta\theta}^\dag$ and
$\hU_{\delta\theta}^\dag \hU_{\delta\theta} = \huno$, (ii) $\Tr(A B) = \Tr(B A)$
(judiciously)  together with  $\hU_{\delta\theta}  \hU_{\delta\theta} =  \huno$,
(iii) the fact  that $\hrho_{\theta_0}$ and $\ln  \hrho_{\theta_0}$ commute, and
(iv) $  \hU_{\delta\theta} \hU_{\delta\theta}  = \huno$  so that  its derivative
vanishes, we obtain
\begin{eqnarray}  
\Tr\left( \hrho_\theta \, \dv{\theta} \ln \hrho_\theta \right) & = &
\Tr\left(  \hU_{\delta\theta}  \,  \hrho_{\theta_0}  \,  \hU_{\delta\theta}^\dag
\dv{\hU_{\delta\theta}^\dag}{\theta}   \,   \ln  \hrho_{\theta_0}   \,
\hU_{\delta\theta}^\dag \right)\nonumber\\
& & +  \Tr\left( \hU_{\delta\theta} \, \hrho_{\theta_0}  \, \ln \hrho_{\theta_0}
\, \dv{\hU_{\delta\theta}^\dag}{\theta} \right)\nonumber\\
&    =   &    \Tr\left(    \ln   \hrho_{\theta_0}    \,   \hrho_{\theta_0}    \,
\hU_{\delta\theta}^\dag           \dv{\hU_{\delta\theta}^\dag}{\theta}
\right)\nonumber\\
&    &   +    \Tr\left(    \hrho_{\theta_0}   \,    \ln   \hrho_{\theta_0}    \,
\dv{\hU_{\delta\theta}^\dag}{\theta}       \,       \hU_{\delta\theta}
\right)\nonumber\\
&  =  &  \Tr\left(  \hrho_{\theta_0}   \,  \ln  \hrho_{\theta_0}  \,  \dv{\left(
  \hU_{\delta\theta}^\dag \hU_{\delta\theta}\right)}{\theta} \right)\nonumber\\
& = & 0\label{eq:secondtermvanishes}
\end{eqnarray}
Consequently, the  second term  of~\eqref{eq:firstderivative} vanishes  and thus
the first derivative of the relative entropy reduces to
\begin{equation}
\label{eq:firstderivativef}
\dv{\theta}  S\left[  \hrho_\theta  \|   \hrho_{\theta_0}  \right]  =  \Tr\left(
\dv{\hrho_\theta}{\theta} \left( \ln  \hrho_\theta - \ln\hrho_{\theta_0} \right)
\right),
\end{equation}
which indeed vanishes at $\theta = \theta_0$.

Now  differentiating  the   expression~\eqref{eq:firstderivativef},  the  second
derivative of the relative entropy writes
\begin{eqnarray}
\dv[2]{\theta} S\left[ \hrho_\theta \| \hrho_{\theta_0} \right] & = &
\Tr\left( \dv[2]{\hrho_\theta}{\theta}  \left(  \ln \hrho_\theta  -  \ln
\hrho_{\theta_0} \right) \right) \nonumber\\
&&  +  \Tr\left(   \dv{\hrho_\theta}{\theta}  \dv{\theta}  \ln
\hrho_\theta \right),\nonumber
\end{eqnarray}
where the first term of the r.h.s. vanishes at $\theta = \theta_0$, so that
\begin{equation}
\label{eq:secondderivative}
\left.  \dv[2]{\theta}   S\left[  \hrho_\theta   \|  \hrho_{\theta_0}
  \right]\right|_{\theta      =      \theta_0}      =      \left.       \Tr\left(
\dv{\hrho_\theta}{\theta}    \dv{\theta}   \ln    \hrho_\theta
\right)\right|_{\theta = \theta_0}.
\end{equation}
Let  $\hC_{\delta\theta}$  be  the   Hermitian  generator  of  the  unitary
transformation $ \hU_{\delta\theta}$, that is,
\begin{equation}
\label{secondtermb}
\hC_{\delta\theta}  =   -  \imath  \,   \dv{\hU^\dag_{\delta\theta}}{\theta}  \,
\hU_{\delta\theta}      =     \imath      \,     \hU_{\delta\theta}^\dag      \,
\dv{\hU_{\delta\theta}}{\theta}.
\end{equation}
Then, we use expression~\eqref{dlnrhodtheta} for $\dv{\theta} \ln \hrho_\theta$,
and similarly for $\dv{\theta} \hrho_\theta$ (just replace $\ln \hrho_\theta$ by
$\hrho_\theta$)  so that,  after some  algebra in  the same  line as  previously
(property of the trace,  unitary property of $\hU_{\delta\theta}$, commutativity
of    $\ln    \hrho_{\theta_0}$    and   $\hrho_{\theta_0}$,    together    with
$\dv{\hU_{\delta\theta}^\dag}{\theta}      \dv{\hU_{\delta\theta}}{\theta}     =
\dv{\hU_{\delta\theta}^\dag}{\theta}       \,        \hU_{\delta\theta}       \,
\hU_{\delta\theta}^\dag \,\dv{\hU_{\delta\theta}}{\theta}$), one obtains
\begin{equation}
\label{eq:2ndtermsecondderivative}
\Tr\left(  \dv{\hrho_\theta}{\theta} \dv{\theta}  \ln  \hrho_\theta \right)  =
\Tr\left( \hrho_{\theta_0} \left[ \hC_{\delta\theta} , \left[ \hC_{\delta\theta}
    , \ln\hrho_{\theta_0} \right] \right] \right)\!.\qquad\qquad
\end{equation}
Taking Eq.~\eqref{eq:2ndtermsecondderivative}  at $\theta  = \theta_0$  where we
define  $\hC_0  =  \hC_{\delta\theta}|_{\theta  = \theta_0}$  and  plugging  the
expression     into~\eqref{eq:secondderivative},      we     finally     achieve
to~\eqref{eq:QFIdef_commutator}.

\section{$\opL_1[\hO]$, $\opL_2[\hO]$ and $\opL_3[\hO]$ in GDSs.}
\label{appB}

Let us start writing $\hA_k= \frac12  (\hL_k  +
\hL_k^\dag)=\Re({\bf l}_k^\dag)\JJJ\hat {\bf x}=\Re({\bf a}_k)^\tp {\bf \hat p}-\Re({\bf b}_k)^\tp {\bf \hat q}$
and $\hB= \frac1{2 \imath}(\hL_k  - \hL_k^\dag)=\Im({\bf l}_k^\dag)\JJ\hat {\bf x}=\Im({\bf a}_k)^\tp {\bf \hat p}-\Im({\bf b}_k)^\tp {\bf \hat q}$ with $\hL_k$ in~\eqref{Hquadratic} and ${\bf l}^\tp _k=({\bf a}^\tp _k,{\bf b}^\tp _k)$. Therefore, using these expressions in~\eqref{eq:L1} we obtain
\begin{align}
\opL_1[\hO]=-\frac{1}{2\hbar}
\sum_{k=1}^K\sum_{l,j=1}^n&
\left(\Re(({\bf a}_k)_{l})\Re(({\bf a}_k)_{j})[\hat p_l,[\hat p_j,\hat O]]\right.\nonumber\\
&-\Re(({\bf b}_k)_{l})\Re(({\bf a}_k)_{j})[\hat q_l,[\hat p_j,\hat O]]\nonumber\\
&-\Re(({\bf a}_k)_{l})\Re(({\bf b}_k)_{j})[\hat p_l,[\hat q_j,\hat O]]\nonumber\\
&\left.+\Re(({\bf b}_k)_{l})\Re(({\bf b}_k)_{j})[\hat q_l,[\hat q_j,\hat O]]\right).\nonumber
\end{align}
Now, using the identity $\Re(zw^*)=\Re(z)\Re(w)+\Im(z)\Im(w)$ we get to the final result,
\begin{align}
\opL_1[\hO]&=\frac{1}{2}\sum_{s=1}^{2n}\sum_{m=1}^{2n}
\left(\hbar\sum_{k=1}^K\Re(({\bf l}_k)_{s}^*({\bf l}_k)_{m})\right)\times\nonumber\\
&\times \frac{-1}{\hbar^2}[(\JJJ{\bf \hat x})_m,[(\JJJ{\bf \hat x})_s,\hat O]]\nonumber\\
&=\frac{1}{2} \sum_{s=1}^{2n}\sum_{m=1}^{2n} \left(\hbar\Re(\GM)\right)_{s,m}(\mathbb{\hat H}[\hat O])_{m,s}\nonumber\\
&=\frac{1}{2} \tr\left(\DD \mathbb{\hat H}[\hat O]\right).
\end{align}
Here, we used the definition in~\eqref{defhatH}, that $\DD = \hbar\Re(\GM)$ and we define  the matrix  notation:
$([\hby ,  [\hby^\tp  ,  \hO]])_{sm} =  [\hy_s , [  \hy_m ,  \hO]]$. 

In the following we will need the identity
\begin{widetext}
\begin{align}
\hat \x^\tp \QQ\hat O\hat \x=&
\sum_{j=1}^n\sum_{l=1}^n \QQ_{jl}\hat x_j\hat O\hat x_l=
\sum_{j=1}^n\sum_{l=1}^n \QQ_{jl}
\left(-[\hat x_l,[\hat x_j,\hat O]]+\{\hat x_l\hat x_j,\hat O\}-\hat x_l\hat O\hat x_j-\imath\hbar \JJJ_{lj} \hat O\right)\nonumber\\
=&-\hbar^2\tr(\QQ\JJJ\hHH {\hat O}\JJJ)+\tr(\QQ\{{\bf x}{\bf x}^T,\hat O\})-\tr(\QQ{\bf x}\hat O{\bf x}^T)-\imath\hbar\tr(\QQ\JJJ)\hat O\nonumber\\
=&-\hbar^2\tr(\QQ\JJJ\hat \HH[\hat O]\JJJ)
+\hbar \tr\left(\QQ\hat \VV\right)\;\hat O+\hbar \;\hat O\;\tr\left(\QQ\hat \VV\right)
-\tr(\QQ\hat \x\hat O\hat \x^\tp),
\label{identityxMOx}
\end{align}
\end{widetext}
where we used the canonical commutation relation $[\hat x_j,\hat x_l]=\imath\hbar (\mathbb{\JJJ})_{j,l}\hat 1$ 
and the identity 
\begin{equation}
\label{identityxxT}
\hat \x\hat \x^\tp=
\hbar \left(\mathbb{\hat V}+\frac{\imath}{2}\JJJ\hat 1\right),
\end{equation}
with $\hat \VV$ the operator matrix $\hat \VV=(\hat \x\hat \x^\tp +(\hat \x\hat \x^\tp )^\tp )/2$.
In the case when $\hat O=\hat 1$ we have,
\begin{align}
\hat \x^\tp \QQ\hat \x=& 2\hbar \tr\left(\QQ\hat \VV\right)-\tr(\QQ\hat \x\hat \x^\tp)\nonumber\\
=&\hbar \tr\left(\QQ\hat \VV\right)-\frac{\imath\hbar}{2}\tr(\QQ\JJJ) \hat 1,
\label{identityxMx}
\end{align}
using~\eqref{identityxxT}. Identities in Eqs.\eqref{identityxMOx} and~\eqref{identityxMx} are valid for arbitrary matrices
$\QQ$, but the interesting case here is when $\QQ$ is anti-symmetric. Indeed, when $\QQ$  is anti-symmetric, because 
$\JJJ\hat \HH[\hat O]\JJJ$ and $\hat \VV$ are symmetric, we have
\begin{eqnarray}
\hat \x^\tp \QQ\hat O\hat \x=&-\tr(\QQ\hat \x\hat O\hat \x^\tp),
\label{PidentityxMOx}\\
\hat \x^\tp \QQ\hat \x=&-\frac{\imath\hbar}{2}\tr(\QQ\JJJ) \hat 1.
\label{PidentityxMx}
\end{eqnarray}

Using the definition for $\opL_2$ in~\eqref{eq:L1},  we have:
\begin{align}
\opL_2[\hat O] & 
=\frac{1}{2\hbar}\sum_{k=1}^K  \frac{1}{2}\{\hat L_k\hat L_k^\dag-\hat L_k^\dag\hat L_k,\hat O\}\nonumber\\
&=
\frac{1}{2\hbar}\sum_{k=1}^K  \frac{1}{2}
\{-\hat {\bf x}^\tp \JJJ{\bf l}_k{\bf l}_k^\dag\JJJ\hat{\bf x}
+\hat{\bf x}^\tp \JJJ {\bf l}_k^*{\bf l}_k^\tp \JJJ \hat{\bf x}
,\hat O\}\nonumber\\
&=
\frac{1}{2\hbar}
\left\{\hat{\bf x}\JJJ\;
\frac{1}{2}
\left(  \sum_{k=1}^K {\bf l}_k^*{\bf l}_k^\tp -
\sum_{k=1}^K {\bf l}_k{\bf l}_k^\dag\right)\JJJ\hat{\bf x}^\tp ,\hat O
\right\}\nonumber\\
&=
\frac{\imath}{2\hbar}
\left\{\hat {\bf x}\JJJ\;
\left( \frac{\GM^*-\GM}{2\imath}\right)\JJJ\hat{\bf x}^\tp ,\hat O
\right\}\nonumber\\
&-\frac{\imath}{2\hbar}
\left\{\hat{\bf x}\JJJ
\CC\JJJ\hat{\bf x}^\tp ,\hat O\right\}\nonumber\\
&=
-\frac{\imath}{2\hbar}
\left\{
-\imath\frac{\hbar}{2} \tr(\JJJ\CC\JJJ^2)\hat 1,\hat O
\right\}=\frac{1}{2}\tr(\JJJ\CC)\hat O,
\label{proveopL2}
\end{align}
where we used  $\CC=(\GM^*-\GM)/2\imath$, that $\JJJ^2=-\II$, and the identity in~\eqref{PidentityxMx}.

For $\opL_3$ in~\eqref{eq:L3} we can write
\begin{align}
\opL_3[\hat O]&=
\frac{1}{2\hbar}\sum_{k=1}^K (\hat L_k\hat O \hat L^\dag_k-\hat L_k^\dag\hat O \hat L_k)\nonumber\\
&
=
\frac{1}{2\hbar}\sum_{k=1}^K 
\hat {\bf x}^\tp \JJJ(-{\bf l}_k{\bf l}_k^\dag\
+{\bf l}_k^*{\bf l}_k^\tp )\JJJ\hat O\hat{\bf x}
\nonumber\\
&=\frac{1}{2\hbar}
\hat {\bf x}^\tp \JJJ\left(-\GM
+\GM^*\right)\JJJ\hat O\hat{\bf x}=
-\frac{\imath}{\hbar}
\hat {\bf x}^\tp \JJJ\CC\JJJ\hat O\hat{\bf x}\nonumber\\
&=
\frac{\imath}{\hbar}\tr(\JJJ\CC\JJJ{\bf x}\hat O{\bf x}^\tp ).
\end{align}
But, $\hat {\bf  x}\hat O\hat {\bf x}^T=-\hat {\bf x}[\hat{\bf  x}^T,\hat O]+ \hat {\bf x}\hat{\bf x}^T\hat O$ where 
we define the matrix notation:
$(\hby ,
[\hby^\tp  , \hO])_{sm} =  \hy_s , [ \hy_m , \hO]$. Therefore,
\begin{align}
{\cal L}_3[\hat O]&=
\frac{\imath}{\hbar}\tr\left(\JJJ\CC\JJJ{\bf x}\hat O{\bf x}^T\right)\nonumber\\
=&-\frac{\imath}{\hbar}\tr\left(\JJJ\CC\JJJ\hat {\bf x}[\hat{\bf  x}^T,\hat O]\right)+
\imath\tr\left(\JJJ\CC\JJJ\hat O\mathbb{\hat V}\right)
-\frac{1}{2}\tr\left(\JJJ\CC\JJJ^2 \hat O\right)\nonumber\\
&=
-\frac{\imath}{\hbar}\tr\left(\CC\JJJ\hat {\bf x}[\hat{\bf  x}^T,\hat O]\JJJ\right)+
\imath\hat O\tr\left(\JJJ\CC\JJJ\mathbb{\hat V}\right)
+\frac{1}{2}\tr\left(\JJJ\CC \right)\hat O\nonumber\\
&=
\tr\left(\CC\JJJ\hMM[\hO]\right)+{\cal L}_{2}[\hat O],
\end{align}
where we define the superoperator matrix $\hMM[\hO]  =  \frac{\imath}{\hbar}(\x)[(\JJJ  \x)^\tp,\hO]$ and we used~\eqref{proveopL2}, that $\tr(\JJJ\CC\JJJ\hat\VV)=0$, $\JJJ^2=-\II$ and
\begin{align}
\label{JxconmuJ}
(\JJJ\hat {\bf x}[\hat{\bf  x}^T,\hat O]\JJJ)_{lj}=&
\sum_{r=1}^{2n}\sum_{s=1}^{2n}\JJJ_{lr}\hat x_r[\hat x_s,\hat O]\JJJ_{sj}\nonumber\\
=&\sum_{r=1}^{2n}\sum_{s=1}^{2n}\JJJ_{lr}\hat x_r[\JJJ_{sj}\hat x_s,\hat O]\nonumber\\
=&-\sum_{r=1}^{2n}\JJJ_{lr}\hat x_r\left[\sum_{s=1}^{2n}\JJJ_{js}\hat x_s,\hat O\right]\nonumber\\
&=-(\JJJ\hat {\bf x}[(\JJJ\hat{\bf  x})^T,\hat O])_{lj}.
\end{align}


\bibliographystyle{apsrev4-2}

%

\end{document}